\documentstyle[myart,12pt]{article}

\renewcommand{\theequation}{\arabic{section}.\arabic{equation}}
\normalbaselines
\font\tenbm=cmmib10
\font\sevenbm=cmmib7
\font\fivebm=cmmib5
\newfam\bmfam
\textfont\bmfam=\tenbm \scriptfont\bmfam=\sevenbm
\scriptscriptfont\bmfam=\fivebm
{\count0=\number\bmfam \multiply\count0 by "100
\def\defbgreek#1#2#3{{\count1=\count0 \advance\count1 by "#2#3
  \global\mathchardef#1=\count1 }}
\defbgreek\balpha  0B \defbgreek\brho       1A
\defbgreek\bbeta   0C \defbgreek\bsigma     1B
\defbgreek\bgamma  0D \defbgreek\btau       1C
\defbgreek\bdelta  0E \defbgreek\bupsilon   1D
\defbgreek\bepsilon0F \defbgreek\bphi       1E
\defbgreek\bzeta   10 \defbgreek\bchi       1F
\defbgreek\bmeta   11 \defbgreek\bpsi       20
\defbgreek\btheta  12 \defbgreek\bomega     21
\defbgreek\biota   13 \defbgreek\bvarepsilon22
\defbgreek\bkappa  14 \defbgreek\bvartheta  23
\defbgreek\blambda 15 \defbgreek\bvarpi     24
\defbgreek\bmu     16 \defbgreek\bvarrho    25
\defbgreek\bnu     17 \defbgreek\bvarsigma  26
        \defbgreek\bxi     18 \defbgreek\bvarphi    27
\defbgreek\bpi     19}
\input tcilatex.tex

\begin{document}

\author{Yuri A. Rylov}
\title{Dynamic disquantization of Dirac equation.}
\date{Institute for Problems in Mechanics, Russian Academy of Sciences \\
101-1 ,Vernadskii Ave., Moscow, 117526, Russia \\
email: rylov@ipmnet.ru\\
Web site: {$http://rsfq1.physics.sunysb.edu/\symbol{126}rylov/yrylov.htm$}\\
or mirror Web site: {$http://194.190.131.172/\symbol{126}rylov/yrylov.htm$}}
\maketitle

\begin{abstract}
Classical model ${\cal S}_{{\rm dc}}$ of Dirac particle ${\cal S}_{{\rm D}}$
is constructed. ${\cal S}_{{\rm D}}$ is the dynamic system described by the
Dirac equation. Its classical analog ${\cal S}_{{\rm dc}}$ is described by a
system of ordinary differential equations, containing the quantum constant $%
\hbar $ as a parameter. Dynamic equations for ${\cal S}_{{\rm dc}}$ are
determined by the Dirac equation uniquely. Both dynamic systems ${\cal S}_{%
{\rm D}}$ and ${\cal S}_{{\rm dc}}$ appear to be nonrelativistic. One
succeeded in transforming nonrelativistic dynamic system ${\cal S}_{{\rm dc}%
} $ into relativistic one ${\cal S}_{{\rm dcr}}$. The dynamic system ${\cal S%
}_{{\rm dcr}}$ appears to be a two-particle structure (special case of a
relativistic rotator). It explains freely such properties of ${\cal S}_{{\rm %
D}}$ as spin and magnetic moment, which are strange for pointlike structure.
The rigidity function $f_{{\rm r}}\left( a\right) $, describing rotational
part of total mass as a function of the radius $a$ of rotator, has been
calculated for ${\cal S}_{{\rm dcr}}$. For investigation of ${\cal S}_{{\rm D%
}}$ and construction of ${\cal S}_{{\rm dc}}$ one uses new dynamic methods:
dynamic quantization and dynamic disquantization. These relativistic pure
dynamic procedures do not use principles of quantum mechanics (QM). They
generalize and replace conventional quantization and transition to classical
approximation. Totality of these methods forms the model conception of
quantum phenomena (MCQP). Technique of MCQP is more subtle and effective,
than conventional methods of QM. MCQP relates to conventional QM, much as
the statistical physics relates to thermodynamics.
\end{abstract}


\newpage

\section{Introduction}

In this investigation we try to answer the question, whether the Dirac
particle has a structure, or it is simply a pointlike particle, which has
such unusual for pointlike particle properties as spin and magnetic moment.
To answer this question, we use procedure of dynamic disquantization which
is a more subtle, than conventional procedure of transition to classical
approximation. Dynamic disquantization (D-disquantization) of dynamic system
${\cal S}_{{\rm D}}$, described by the Dirac equations, results in classical
relativistic rotator ${\cal S}_{{\rm rr}}$, consisting of two coupled
charged particles, rotating around their mass center. The more rough
classical approximation, being applied to ${\cal S}_{{\rm D}}$, results in
pointlike classical particle, possessing spin and magnetic moment. A
characteristic of any rotator is the rigidity function $f_{{\rm r}}$ which
is defined as follows. Let $E_{{\rm tot}}\left( a\right) $ be total rest
energy of the rotator as a function of the radius $a$. The rigidity function
$f_{{\rm r}}$ connects the relative rotational energy $\gamma =E_{{\rm rot}%
}/E_{{\rm tot}}\left( 0\right) $ with the radius $a$ of the rotator by means
of relation
\begin{equation}
\gamma =f_{{\rm r}}\left( a\right) .  \label{g1.1}
\end{equation}
For nonrelativistic rotator ${\cal S}_{{\rm nr}}$ the rigidity function $f_{%
{\rm r}}$ is connected with the potential energy $U\left( a\right) $ of the
coupling by means of relation
\[
f_{{\rm r}}\left( a\right) =\frac{a}{m_{0}c^{2}}\frac{\partial U\left(
a\right) }{\partial a}
\]
where $m_{0}$ is the mass of a rotator particle. The potential energy $U$ of
the coupling cannot be introduced for a relativistic rotator, but the
rigidity function $f_{{\rm r}}$ of the rotator coupling may be introduced in
any case.

The rigidity function for the rotator ${\cal S}_{{\rm rr}}$ which is a
classical analog of Dirac dynamic system ${\cal S}_{{\rm D}}$ appears to
have the form
\begin{equation}
\gamma =f_{{\rm r}}\left( a\right) =\frac{\hbar }{\sqrt{\hbar ^{2}-\left(
4am_{0}c\right) ^{2}}}-1  \label{b8.80}
\end{equation}
One can see that in (\ref{b8.80}) $\hbar >4am_{0}c$, and one can set $\hbar
\rightarrow 0$ only in the case, if simultaneously $a\rightarrow 0$. The
magnetic moment and spin are natural properties of any charged rotator.

Conventional disquantization (classical approximation) of the Dirac equation
cannot give the result (\ref{b8.80}), because it sets $\hbar =0$ in the
final result. It obtains only a charged pointlike particle with magnetic
moment and spin. In general, a classical analog, obtained as a classical
approximation of a quantum system, cannot contain the quantum constant $%
\hbar $.

The more detailed information on classical origin of a quantum system is
obtained, because one uses the model conception of quantum phenomena\
(MCQP), which relates to the conventional (axiomatic) quantum theory in much
the same way, that the statistical physics (model conception of thermal
phenomena) relates to the axiomatic thermodynamics. Ignoring the internal
degrees of freedom of ${\cal S}_{{\rm D}}$ is justified, when one calculates
atom structures. In this case the interaction energy is low, internal
degrees of freedom are not excited, and one can neglect them. If one
investigates the structure of elementary particles, where the interaction
energies are very high, one should use model conception of quantum phenomena
(MCQP), which enables one to take into account details of internal
structure. Investigation of elementary particles by means of conventional
axiomatic quantum theory resembles investigation of crystal structure by
means of thermodynamics without having an idea on molecular structure of the
matter.

The conventional disquantization (Q-disquantization) of quantum dynamic
system ${\cal S}_{{\rm q}}$ results in derivation of classical analog ${\cal %
S}_{{\rm c}}$ of quantum system ${\cal S}_{{\rm q}}$. The dynamic system $%
{\cal S}_{{\rm c}}$ has two properties:

(1) Dynamic equations for ${\cal S}_{{\rm c}}$ are ordinary differential
equations.

(2) Dynamic equations for ${\cal S}_{{\rm c}}$ do not contain quantum
constant $\hbar $.

It is common practice to think that the two properties are connected
closely, and there are no such ${\cal S}_{{\rm c}}$, whose dynamic equations
be ordinary dynamic equations, containing $\hbar $ as a parameter. Model
conception of quantum phenomena (MCQP) separates the two properties of $%
{\cal S}_{{\rm c}}$. In the scope of MCQP there exist such classical models $%
{\cal S}_{{\rm c}}$ of quantum system ${\cal S}_{{\rm q}}$, where dynamic
equations for ${\cal S}_{{\rm c}}$ are ordinary differential equations,
containing $\hbar $ as a parameter. Of course, one can set in this model $%
\hbar =0$ and derive a more rough model, which is obtained as a result of
conventional disquantization (Q-disquantization). Dynamic models of MCQP are
intermediate between pure quantum models and pure classical ones.
Practically, models of MCQP are classical models, containing parameter $%
\hbar ,$ and these models are more subtle, than conventional classical
models.

Let us note that the dynamic disquantization (D-disquantization) is simply a
{\it method of investigation} of  dynamic system ${\cal S}_{{\rm D}}$, when
the complicated dynamic system ${\cal S}_{{\rm D}}$, described by a system
of partial differential equations, associates with a more simple dynamic
system ${\cal S}_{{\rm c}}$, described by a system of ordinary differential
equations.

For instance, let ${\cal S}_{{\rm KG}}$ be dynamic system described by the
free Klein-Gordon equation. Conventional classical analog ${\cal S}_{{\rm KGc%
}}$ of ${\cal S}_{{\rm KG}}$ is spinless relativistic particle. In the scope
of MCQP the classical analog ${\cal S}_{{\rm KGqu}}$ of ${\cal S}_{{\rm KG}}$
is a classical relativistic particle (described by ordinary differential
equations) in some force field $\kappa ,$ generated by this particle \cite
{R98}. The field $\kappa $ describes a cloud of pairs around the particle.
Interaction between the particle and the field $\kappa $ contains the
parameter $\hbar ,$ and vanishes, if one sets $\hbar =0$.

The conventional quantum theory states that the classical relativistic
particle cannot produce pairs, but quantum one can. Why? Nobody knows. MCQP
states that origin of pair production is a special force field, and
quantization in itself has nothing to do with pair production.

MCQP uses special procedures: dynamic quantization (D-quantization) and
dynamic disquantization (D-disquantization), which are dynamic
generalizations of conventional quantum procedures of quantization
(Q-quantization) and of the classical approximation derivation
(Q-disquantization). Mathematical ground of MCQP is three different methods,
used for description of dynamic systems, known as statistical ensembles.

Let ${\cal S}_{{\rm c}}$ be classical dynamic system described by the action
\begin{equation}
{\cal S}_{{\rm c}}:\qquad {\cal A}_{{\rm c}}\left[ {\bf x}\right] =\int
L\left( t,{\bf x},{\bf \dot{x}}\right) dt  \label{g1.2}
\end{equation}
Here $L\left( t,{\bf x},{\bf \dot{x}}\right) =L\left( x,{\bf v}\right) $ is
its Lagrangian, and ${\bf x}=\left\{ x^{\alpha }\right\} ,$ \ \ ${\bf v}=%
{\bf \dot{x}}=\left\{ \frac{dx^{\alpha }}{dt}\right\} $,$\;\;\alpha
=1,2,...n.$ The statistical ensemble ${\cal E}\left[ {\cal S}_{{\rm c}}%
\right] $, consisting of ${\cal S}_{{\rm c}}$, is described by the action
\begin{equation}
{\cal E}\left[ {\cal S}_{{\rm c}}\right] :\qquad {\cal A}_{{\rm E}}\left[
{\bf x}\right] {\bf =}\int {\cal A}_{{\rm c}}\left[ {\bf x}\right] d{\bf %
\tau }=\int L\left( t,{\bf x},{\bf \dot{x}}\right) dtd{\bf \tau }
\label{g1.3}
\end{equation}
where ${\bf x}={\bf x}\left( t,{\bf \tau }\right) $ are considered to be
functions of Lagrangian coordinates $\tau =\left\{ t,{\bf \tau }\right\} ,$
\ \ ${\bf \tau =}\left\{ \tau _{\alpha }\right\} ,$ \ \ $\alpha =1,2,...n$,
which label elements (dynamic systems ${\cal S}_{{\rm c}}$) of the
statistical ensemble ${\cal E}\left[ {\cal S}_{{\rm c}}\right] .$ This
labeling is arbitrary, and the action (\ref{g1.3}) is invariant with respect
to relabeling of elements ${\cal S}_{{\rm c}}$
\begin{equation}
\tau _{\alpha }\rightarrow \tilde{\tau}_{\alpha }=\tilde{\tau}_{\alpha
}\left( {\bf \tau }\right) ,\qquad \alpha =1,2,...n,\qquad \frac{\partial
\left( \tilde{\tau}_{1},\tilde{\tau}_{2},...\tilde{\tau}_{n},\right) }{%
\partial \left( \tau _{1},\tau _{2},...\tau _{n},\right) }=1  \label{g1.4}
\end{equation}
This method of description will be referred to as L-description, or
L-representation, because it uses Lagrangian coordinates ${\bf \tau }$ as
independent variables.

The statistical ensemble ${\cal E}\left[ {\cal S}_{{\rm c}}\right] $ is a
kind of a fluid without pressure. The action (\ref{g1.3}) can be rewritten
in the space of variables $x^{i}=\left\{ t,{\bf x}\right\} =\left\{ x^{0},%
{\bf x}\right\} $ in terms of hydrodynamic variables: flux n-vector $%
j^{i}=\left\{ \rho ,{\bf j}\right\} =\left\{ \rho ,\rho {\bf v}\right\} $
and hydrodynamic potentials (Clebsch potentials \cite{C57,C59}) $\tau
=\left\{ \varphi ,{\bf \tau }\right\} $. The action (\ref{g1.3}) transforms
to the form
\begin{equation}
{\cal E}\left[ {\cal S}_{{\rm c}}\right] :\qquad {\cal A}_{{\rm E}}[\rho ,%
{\bf v},\varphi ,{\bf \tau }]=\int \rho \{L(x,{\bf v})-{\bf vp}-p_{0}\}{\rm d%
}^{n+1}x,  \label{g1.5}
\end{equation}
Here dependent variables $\rho ,{\bf v},\varphi ,{\bf \tau }$ are considered
to be functions of independent variables $x$. The quantities $p_{k}=\left\{
p_{0},{\bf p}\right\} =\left\{ p_{k}\right\} ,\;\;k=0,1,...n$ are defined by
the relations
\begin{equation}
p_{k}=b_{0}\left( \partial _{k}\varphi +g^{\alpha }\left( {\bf \tau }\right)
\partial _{k}\tau _{\alpha }\right) ,\qquad k=0,1,...n  \label{g1.6}
\end{equation}
where $g^{a}\left( {\bf \tau }\right) ,\;\;\alpha =1,2,...n$ are arbitrary
functions of only ${\bf \tau }$, and $b_{0}$ is an arbitrary constant. The
action (\ref{g1.5}) is obtained from the action (\ref{g1.3}) by a change of
variables. Now the variables $x$ are independent variables, and variables $%
{\bf \tau }$ are dependent ones. Besides one introduces designation
\begin{equation}
j^{i}=\left\{ \rho ,\rho {\bf v}\right\} =\frac{\partial (x^{i},\tau
_{1},...\tau _{n})}{\partial (x^{0},x^{1},...x^{n})},\qquad i=0,1,...n
\label{g1.6a}
\end{equation}
The change of variables is accompanied by integration of some dynamic
equations. The arbitrary functions $g^{a}\left( {\bf \tau }\right)
,\;\;\alpha =1,2,...n$ and the arbitrary constant $b_{0}$ are results of
this integration. This integration appears to be possible, because the
dynamic system ${\cal E}\left[ {\cal S}_{{\rm c}}\right] $ has the symmetry
group (\ref{g1.4}), containing $n$ arbitrary functions. We shall not go into
details of this change of variables, which is rather complicated. One can
find these details in the paper \cite{R99}. This method, using Lagrangian
coordinates ${\bf \tau }$ as hydrodynamic potentials, will be referred to as
HDP-description, or HDP-representation.

The action (\ref{g1.5}) can be described also in terms of $\psi $ function
(wave function). Let us introduce $k$-component complex function $\psi
=\{\psi _{\alpha }\},\;\;\alpha =1,2,\ldots k$, defining it by the relations

\begin{equation}
\psi _{\alpha }=\sqrt{\rho }e^{i\varphi }u_{\alpha }({\bf \tau }),\qquad
\psi _{\alpha }^{\ast }=\sqrt{\rho }e^{-i\varphi }u_{\alpha }^{\ast }({\bf %
\tau }),\qquad \alpha =1,2,\ldots k  \label{g1.7}
\end{equation}
\begin{equation}
\psi ^{\ast }\psi \equiv \sum_{\alpha =1}^{k}\psi _{\alpha }^{\ast }\psi
_{\alpha }  \label{g1.8}
\end{equation}
where (*) means the complex conjugate, $u_{\alpha }({\bf \tau })$, $\;\alpha
=1,2,\ldots k$ are functions of only variables ${\bf \tau }$. They satisfy
the relations
\begin{equation}
-\frac{i}{2}\sum_{\alpha =1}^{k}(u_{\alpha }^{\ast }\frac{\partial u_{\alpha
}}{\partial \tau _{\beta }}-\frac{\partial u_{\alpha }^{\ast }}{\partial
\tau _{\beta }}u_{\alpha })=g^{\beta }({\bf \tau }),\qquad \beta
=1,2,...n,\qquad \sum_{\alpha =1}^{k}u_{\alpha }^{\ast }u_{\alpha }=1
\label{g1.9}
\end{equation}
$k$ is such a natural number that equations (\ref{g1.9}) admit a solution.
In general, the minimal possible value of $k$ depends on the form of the
arbitrary integration functions ${\bf g}=\{g^{\beta }({\bf \tau })\}$,\ \ $%
\beta =1,2,...n.$ After such a change of variables the action (\ref{g1.5})
takes the form
\begin{equation}
{\cal E}\left[ {\cal S}_{{\rm c}}\right] :\qquad {\cal A}_{{\rm E}}[\psi
,\psi ^{\ast }]=\int \rho \{L(x,{\bf V})-{\bf VP}-P_{0}\}{\rm d}^{n+1}x,
\label{g1.10}
\end{equation}
where
\begin{equation}
\rho =\psi ^{\ast }\psi ,\qquad P_{l}=P_{l}(\psi ,\psi ^{\ast })=-\frac{%
ib_{0}}{2\psi ^{\ast }\psi }(\psi ^{\ast }\partial _{l}\psi -\partial
_{l}\psi ^{\ast }\cdot \psi ),\qquad l=0,1,...n  \label{g1.11}
\end{equation}
and ${\bf V=}\left\{ V^{\alpha }\right\} ,\;\;\alpha =1,2,...n$ is some
function of ${\bf P}$ and $x$, defined by the equations
\begin{equation}
\frac{\partial L(x,{\bf V})}{\partial V^{\beta }}-P_{\beta }=0,\qquad \beta
=1,2,...n  \label{g1.12}
\end{equation}
If dynamic system ${\cal S}_{{\rm c}}$ has the Hamilton function
\begin{equation}
H\left( x,{\bf p}\right) ={\bf pv-L}\left( x,{\bf v}\right)  \label{g1.13}
\end{equation}
the action (\ref{g1.10}) can be written in the form
\begin{equation}
{\cal E}\left[ {\cal S}_{{\rm c}}\right] :\qquad {\cal A}_{{\rm E}}[\psi ,%
\bar{\psi}]=\int \rho \{-H(x,{\bf P})-P_{0}\}{\rm d}^{n+1}x,  \label{g1.14}
\end{equation}
where $P_{k}$ and $\rho $ are defined by relations (\ref{g1.11}). This
method \cite{R99}, describing the dynamic system ${\cal E}\left[ {\cal S}_{%
{\rm c}}\right] $ in terms of wave function, will be referred to as
WF-description, or WF-representation.

The three different actions (\ref{g1.3}), (\ref{g1.5}), (\ref{g1.14}) can be
also used for description of dynamic system ${\cal S}\left[ {\cal S}_{{\rm c}%
}\right] $ which is a set of interacting classical dynamic systems ${\cal S}%
_{{\rm c}}$. In this case the actions (\ref{g1.3}), (\ref{g1.5}), (\ref
{g1.14}) contain additional term ${\cal A}_{{\rm int}}$, describing
interaction of dynamic systems ${\cal S}_{{\rm c}}$.

Application of three different methods of description of the same dynamic
system is useful in the relation, that this admits one to distinguish
between the properties of dynamic system and those of the description
method. It is very important for investigation of the dynamic system
properties. All properties, that are common for the three methods of
descriptions attribute to dynamic system in itself. If one uses only one
method of description, for instance, WF-description, it is rather difficult
to distinguish between what attributes to the wave function properties and
what attributes to the properties of the dynamic system investigated. For
instance, it is common practice to think that the wave function is an
attribute of quantum dynamic system, and that the wave function is a
fundamental concept, connected with quantum systems, whereas the wave
function is only a method of description, which can be applied to any
dynamic system (classical or quantum). Use of three different methods of
description is useful also in the sense, that L-description is best suited
for interpretation. The HDP-description is best suited to realize the
D-disquantization and to derive classical analog. The WF-description is best
suited to solve dynamic equations, because sometimes they are linear in
WF-representation.

Let us list the properties of the three different descriptions in
application to the statistical ensemble ${\cal E}\left[ {\cal S}_{{\rm c}}%
\right] $.

First, the three representations of the action (\ref{g1.3}), (\ref{g1.5})
and (\ref{g1.10}) for ${\cal E}\left[ {\cal S}_{{\rm c}}\right] $ contain
derivatives only in one direction. The action (\ref{g1.3}) contains only
derivatives with respect to $t,$ the action (\ref{g1.5}) contains
derivatives only in the direction of the flux $n$-vector $j^{i}=\left\{ \rho
,\rho {\bf v}\right\} .$
\begin{equation}
-\rho \left( {\bf vp+}p_{0}\right) =b_{0}j^{k}\left( \partial _{k}\varphi
+g^{\alpha }\left( {\bf \tau }\right) \right) \partial _{k}\tau _{\alpha }
\label{g1.15}
\end{equation}
In the action (\ref{g1.14}) this fact also takes place, but it is not
evident, because it does not contain the variables $j^{i}$ explicitly.

Second, the state of the ensemble ${\cal E}\left[ {\cal S}_{{\rm c}}\right] $
is determined only within the transformation (\ref{g1.4}). The relabeling
group is very powerful, and ambiguity of description is very large. It is
especially valid for the wave function $\psi $, whose meaning is very
obscure. Obscurity of the wave function $\psi $ is greater, than that of
hydrodynamic potentials ${\bf \tau },$ because the arbitrary functions $%
g^{\alpha }$ are ''hidden'' inside the wave function $\psi $. To simplify
the WF-description, one can impose some constraints on the choice of the
wave function. Conventionally one demands that the wave function would
satisfy a linear dynamic equation. It is not clear whether such a constraint
could be imposed always, but in many physically interesting cases this is
possible. Such constraints simplify dealing with wave functions.
Nevertheless, meaning of the wave function remains obscure. To work with
wave functions, considered to be fundamental objects of theory, one has
worked out a system of rules (quantum principles).

Third, the properties of the statistical ensemble ${\cal E}\left[ {\cal S}_{%
{\rm c}}\right] $ do not depend on the number of elements ${\cal S}_{{\rm c}%
} $ in ${\cal E}\left[ {\cal S}_{{\rm c}}\right] $. The actions (\ref{g1.5}%
), (\ref{g1.10}) have respectively the following property
\begin{equation}
{\cal A}_{{\rm E}}[\alpha \rho ,{\bf v},\varphi ,{\bf \tau }]=\alpha {\cal A}%
_{{\rm E}}[\rho ,{\bf v},\varphi ,{\bf \tau }],\qquad {\cal A}_{{\rm E}}[%
\sqrt{\alpha }\psi ,\sqrt{\alpha }\psi ^{\ast }]=\alpha {\cal A}_{{\rm E}%
}[\psi ,\psi ^{\ast }]  \label{g1.16}
\end{equation}
where $\alpha $ is an arbitrary positive constant.

Dynamic quantization (D-quantization) is an addition of a term ${\cal A}_{%
{\rm int}}$ to the action ${\cal A}_{{\rm E}}$ for ${\cal E}\left[ {\cal S}_{%
{\rm c}}\right] .$ The term ${\cal A}_{{\rm int}}$ contains transversal
derivatives $\partial _{\perp }^{l}$, defined by relations
\begin{equation}
\partial _{\parallel }^{k}\equiv \frac{j^{k}j^{s}}{\rho ^{2}}\partial
_{s},\qquad \partial _{\perp }^{k}\equiv \partial ^{k}-\frac{j^{k}j^{s}}{%
\rho ^{2}}\partial _{s},\qquad \partial ^{k}=\partial _{\perp }^{k}+\partial
_{\parallel }^{k},\qquad \rho ^{2}\equiv j^{s}j_{s}  \label{f1.4}
\end{equation}
The transversal derivatives $\partial _{\perp }^{l}$ describe interaction
between the elements ${\cal S}_{{\rm c}}$ of the statistical ensemble ${\cal %
E}\left[ {\cal S}_{{\rm c}}\right] .$ Dynamic equations for ${\cal E}\left[
{\cal S}_{{\rm c}}\right] $ stop to be ordinary differential equations and
form a system of partial differential equations. As a consequence of
D-quantization the ensemble ${\cal E}\left[ {\cal S}_{{\rm c}}\right] $
turns to ${\cal E}_{{\rm red}}\left[ {\cal S}_{{\rm c}}\right] \equiv {\cal S%
}_{{\rm q}}\left[ {\cal S}_{{\rm c}}\right] $ which describes a set of
interacting dynamic systems ${\cal S}_{{\rm c}}$. From formal viewpoint the
D-quantization is a very general dynamic procedure, which do not contain any
conventional attributes of quantization such as linear differential
equations, quantum constant $\hbar $, linear operators etc. Nevertheless,
under some conditions the D-quantization leads to he same results, as the
conventional Q-quantization.

The dynamic system ${\cal E}_{{\rm red}}\left[ {\cal S}_{{\rm c}}\right]
\equiv {\cal S}_{{\rm q}}\left[ {\cal S}_{{\rm c}}\right] $ remains to be a
statistical ensemble, because it continues to satisfy the condition (\ref
{g1.16}). The dynamic system ${\cal S}_{{\rm q}}\left[ {\cal S}_{{\rm c}}%
\right] $ may be considered to be a set of similar independent stochastic
systems ${\cal S}_{{\rm st}}$, i.e. ${\cal S}_{{\rm q}}\left[ {\cal S}_{{\rm %
c}}\right] \equiv {\cal E}\left[ {\cal S}_{{\rm st}}\right] .$ There is no
dynamic equations for a single ${\cal S}_{{\rm st}},$ but dynamic equations
exist for statistical ensemble ${\cal S}_{{\rm q}}\left[ {\cal S}_{{\rm c}}%
\right] \equiv {\cal E}\left[ {\cal S}_{{\rm st}}\right] ,$ and
investigating ${\cal E}\left[ {\cal S}_{{\rm st}}\right] ,$ one studies the
stochastic systems ${\cal S}_{{\rm st}}$. In fact, investigation of dynamic
properties of the statistical ensemble ${\cal E}\left[ {\cal S}_{{\rm st}}%
\right] $ is the only way of investigation of stochastic systems ${\cal S}_{%
{\rm st}}$. One can study distributions of different physical quantities and
their evolution in the statistical ensemble ${\cal E}\left[ {\cal S}_{{\rm st%
}}\right] ,$ and make some conclusions about properties of ${\cal S}_{{\rm st%
}}$ on this base.

Let us show properties of D-quantization in the example of the free
classical particle, which is described by the Lagrangian $L$ and Hamiltonian
$H,$ defined by
\begin{equation}
L\left( x,{\bf v}\right) =-mc^{2}+\frac{m{\bf v}^{2}}{2},\qquad H\left( x,%
{\bf p}\right) =mc^{2}+\frac{{\bf p}^{2}}{2m}  \label{g1.17}
\end{equation}
The interaction term ${\cal A}_{{\rm int}}$ is supposed to have the form
\begin{equation}
{\cal A}_{{\rm int}}[\rho ,{\bf v},\varphi ,{\bf \tau }]=-\int \rho \frac{%
{\bf p}_{{\rm st}}^{2}}{2m}{\rm d}^{n+1}x,\qquad {\bf p}_{{\rm st}}=-\frac{%
\hbar }{2}{\bf \nabla }\ln \rho  \label{g1.18}
\end{equation}
Formally this term coincides with the Bohm potential energy \cite{B52}.
Interpretation of this term is as follows. The motion of free particles is
supposed to be stochastic, and ${\bf p}_{{\rm st}}=-\frac{\hbar }{2}{\bf %
\nabla }\ln \rho $ describes the mean momentum of the stochastic motion,
where $\rho $ is a collective variable
\begin{equation}
\rho =\frac{\partial \left( \tau _{1},\tau _{2},\tau _{3}\right) }{\partial
\left( x^{1},x^{2},x^{3}\right) },  \label{g1.18a}
\end{equation}
describing density of elements ${\cal S}_{{\rm c}}$ in the ensemble ${\cal E}%
\left[ {\cal S}_{{\rm c}}\right] $. To take into account the stochasticity
influence on the regular motion, the energy of the mean stochastic motion,
described by (\ref{g1.18}), should be added to the action (\ref{g1.5}). Let
us consider for simplicity the case, when the flow in ${\cal S}_{{\rm q}}%
\left[ {\cal S}_{{\rm c}}\right] $ is irrotational and one can set $%
g^{\alpha }=0,\;\;\alpha =1,2,3$ \cite{R99}. In this case the wave function $%
\psi $ can have one component. Then the actions (\ref{g1.3}), (\ref{g1.5})
and (\ref{g1.10}) for ${\cal E}\left[ {\cal S}_{{\rm c}}\right] $ take
respectively the form
\begin{equation}
{\cal S}_{{\rm q}}\left[ {\cal S}_{{\rm c}}\right] :\qquad {\cal A}_{{\rm q}}%
\left[ {\bf x},{\bf u}_{{\rm st}}\right] {\bf =}\int \left\{ -mc^{2}+\frac{m%
{\bf \dot{x}}^{2}}{2}+\frac{m{\bf u}_{{\rm st}}^{2}}{2}-\frac{\hbar }{2}{\bf %
\nabla u}_{{\rm st}}\right\} dtd{\bf \tau }  \label{g1.19}
\end{equation}
where ${\bf x}={\bf x}\left( t,{\bf \tau }\right) $ and ${\bf u}_{{\rm st}}=%
{\bf u}_{{\rm st}}\left( t,{\bf x}\right) $ is a new dynamic variable,
describing stochastic component of motion.
\begin{equation}
{\cal S}_{{\rm q}}\left[ {\cal S}_{{\rm c}}\right] :\qquad {\cal A}_{{\rm q}%
}[\rho ,\varphi ]=\int \rho \{-mc^{2}-\frac{b_{0}^{2}\left( {\bf \nabla }%
\varphi \right) ^{2}}{2m}-\frac{\hbar ^{2}}{8m}\left( {\bf \nabla }\ln \rho
\right) ^{2}-b_{0}\partial _{0}\varphi \}{\rm d}^{4}x,  \label{b3.11}
\end{equation}
\[
{\cal S}_{{\rm q}}\left[ {\cal S}_{{\rm c}}\right] :\qquad {\cal A}_{{\rm q}%
}[\psi ,\psi ^{\ast }]=\int \{\frac{ib_{0}}{2}(\psi ^{\ast }\partial
_{0}\psi -\partial _{0}\psi ^{\ast }\cdot \psi )-\frac{b_{0}^{2}}{2m}{\bf %
\nabla }\psi ^{\ast }\cdot {\bf \nabla }\psi
\]
\begin{equation}
-mc^{2}\rho +\frac{b_{0}^{2}-\hbar }{8\rho m}^{2}(\nabla \rho )^{2}\}{\rm d}%
^{4}x,\qquad \rho \equiv \psi ^{\ast }\psi  \label{b3.13}
\end{equation}

Interaction between the elements ${\cal S}_{{\rm c}}$ of the reduced
ensemble ${\cal S}_{{\rm q}}\left[ {\cal S}_{{\rm c}}\right] $ is described
in actions (\ref{b3.11}), (\ref{b3.13}) by means of terms at the factor $%
\hbar ^{2}$. In the action (\ref{g1.19}) interaction is described by means
of additional dependent dynamic variable ${\bf u}_{{\rm st}}$, which depends
on variables $\left\{ t,{\bf x}\right\} {\bf ,}$ but not on $\left\{ t,{\bf %
\tau }\right\} {\bf ,}$ and varying the action with respect to ${\bf u}_{%
{\rm st}}$, one has to use $\left\{ t,{\bf x}\right\} $ as independent
variables
\[
\int (\frac{m{\bf u}_{{\rm st}}^{2}}{2}-\frac{\hbar}2 {\bf \nabla u}_{{\rm st%
}})dtd{\bf \tau \rightarrow }\int (\frac{m{\bf u}_{{\rm st}}^{2}}{2}-\frac{%
\hbar }{2}{\bf \nabla u}_{{\rm st}})\rho dtd{\bf x,\qquad }\rho \equiv \frac{%
\partial \left( \tau _{1},\tau _{2},\tau _{3}\right) }{\partial \left(
x^{1},x^{2},x^{3}\right) }
\]
Then variation with respect to ${\bf u}_{{\rm st}}$ leads to the dynamic
equation
\begin{equation}
m\rho {\bf u}_{{\rm st}}{\bf +}\frac{\hbar }{2}{\bf \nabla }\rho =0,\qquad
{\bf u}_{{\rm st}}{\bf =-}\frac{\hbar }{2m}{\bf \nabla }\ln \rho
\label{g1.20}
\end{equation}
which is equivalent to (\ref{g1.18}). Thus, the term $m{\bf u}_{{\rm st}%
}^{2}/2$ in (\ref{g1.18}) describes the energy of stochastic component of
motion, whereas the term $-\hbar {\bf \nabla u}_{{\rm st}}/2$ describes
interaction of the stochastic component with regular one.

The action (\ref{b3.13}) of WF-description generates, in general, nonlinear
dynamic equation for $\psi $. But if $b_{0}^{2}=\hbar ^{2}$, the nonlinear
term vanishes, and the dynamic equation becomes linear. The constant $b_{0}$
is an arbitrary constant, and nothing prevents from setting $b_{0}=\hbar $.

Let us set $b_{0}=\hbar .$ The action (\ref{b3.13}) transforms to the action
\begin{equation}
{\cal A}_{{\rm q}}[\psi ,\psi ^{\ast }]=\int \{\frac{i\hbar }{2}(\psi ^{\ast
}\partial _{0}\psi -\partial _{0}\psi ^{\ast }\cdot \psi )-\frac{\hbar ^{2}}{%
2m}{\bf \nabla }\psi ^{\ast }\cdot {\bf \nabla }\psi -mc^{2}\psi ^{\ast
}\psi \}{\rm d}^{4}x  \label{g1.21}
\end{equation}
which generates linear dynamic equation
\begin{equation}
i\hbar \partial _{0}\psi +\frac{\hbar ^{2}}{2m}{\bf \nabla }^{2}\psi
-mc^{2}\psi =0  \label{g1.22}
\end{equation}
The equation (\ref{g1.22}) turns to conventional Schr\"{o}dinger equation
after transformation $\psi \rightarrow \psi \exp \left( -imc^{2}t/\hbar
\right) $. Equivalency of relations (\ref{b3.11}) with $b_{0}=\hbar $ and (%
\ref{g1.21}) has been known for a long time \cite{M26,B26}.

Let us compare the actions (\ref{b3.13}) and (\ref{g1.21}). The action (\ref
{b3.13}) contains only one quantum term, i.e. the term, containing quantum
constant $\hbar .$ If one sets $\hbar =0$, the action (\ref{b3.13}) turns to
the action of the form (\ref{g1.14}), describing noninteracting particles $%
{\cal S}_{{\rm c}}$. In the action (\ref{g1.21}) all terms (except for the
last one) are quantum. If one sets $\hbar =0$ in the action (\ref{g1.21}),
it begin to describe nothing, because all terms except for the last one
vanish. To obtain the classical description from the action (\ref{g1.22}),
one is to apply subtle methods of classical approximation.

The quantum constant $\hbar $ plays different role in conventional quantum
mechanics and in MCQP. In QM the quantum constant $\hbar $ is an attribute
of Q-quantization procedure in the sense that $\hbar $ appears together with
Q-quantization and disappears together with Q-disquantization. In MCQP
D-quantization and D-disquantization do not depend on $\hbar $ directly,
although D-quantization may contain $\hbar $ as a parameter. In MCQP the
constant $\hbar $ is a characteristic of some stochastic agent, which is an
origin of stochastic behavior of microparticles. The constant $\hbar $ as
well as the speed of the light $c$ may be considered to be some
characteristic of space-time \cite{R91,R995}.

By definition D-quantization is an addition of derivatives transversal to
the flux vector $j^{k}$, D-disquantization is elimination of derivatives
transversal to $j^{k}.$ Both procedures are possible only in
HDP-representation, where the flux vector $j^{k}$ is an explicit dependent
variable. In the relativistic case the Q-quantization may introduce
additional degrees of freedom, containing both longitudinal and transversal
derivatives. This circumstance is discovered at the D-disquantization.

The Klein-Gordon equation is the dynamic equation for dynamic system ${\cal S%
}_{{\rm KG}}$. It is written usually in WF-representation. Being written in
L-representation, the action for the dynamic system ${\cal S}_{{\rm KG}}$
has the form \cite{R98}
\begin{equation}
{\cal S}_{{\rm KG}}:\qquad {\cal A}_{{\rm L}}\left[ x,\kappa \right] =\int
-mcK\sqrt{\dot{x}^{k}\dot{x}_{k}}d\tau _{0}d{\bf \tau ,\qquad }K{\bf =}\sqrt{%
1+\lambda ^{2}\left( \kappa _{l}\kappa ^{l}+\partial _{l}\kappa ^{l}\right) }
\label{g1.23}
\end{equation}
where $\kappa ^{l}=\kappa ^{l}\left( x\right) ,\;$ $l=0,1,2,3$ are functions
of argument $x=\left\{ x^{k}\left( \tau _{0},{\bf \tau }\right) \right\} ,$
\ $k=0,1,2,3,$ $\lambda =\hbar /mc$ is the Compton wavelength. The action (%
\ref{g1.23}) is a relativistic generalization of the relation (\ref{g1.19}).
One can verify this, expanding expression for $K$ over degrees of $\lambda
^{2},$ neglecting time component of $\kappa ^{l}$ and setting ${\bf u}_{{\rm %
st}}{\bf =}\hbar {\bf \kappa }/m$. The field $\kappa $ in (\ref{g1.23}) as
well as the mean stochastic velocity ${\bf u}_{{\rm st}}$ in (\ref{g1.19})
describes stochastic agent, which is an origin of stochastic behavior of
microparticles. But there is an essential difference in the two
descriptions. Description in terms of ${\bf u}_{{\rm st}}$ does not contain
time derivatives, and the mean stochastic velocity ${\bf u}_{{\rm st}}$
describes some field, coupled rigidly with its source (particle).
Description in terms of $\kappa $ contains time derivatives, and the field $%
\kappa $ can exist independently of its source (particle). Both fields ${\bf %
u}_{{\rm st}}$ and $\kappa $ describe cloud of pairs around the particle.
The cloud, described by ${\bf u}_{{\rm st}}$, cannot move away from the
particle, whereas the pair cloud, described by the field $\kappa $, can. In
the case (\ref{g1.19}) one can unite the particle and cloud of pairs,
described by ${\bf u}_{{\rm st}}$ into one pointlike object and consider it
to be a clothed particle. In the case (\ref{g1.23}) one cannot join particle
with the cloud of pairs, because the cloud can exist separately from the
particle, taking a part of its energy.

Formally all this follows from dynamic equation for the field $\kappa ,$
which has the form
\begin{equation}
\kappa _{l}=-\frac{1}{2}\partial _{l}\ln \rho ,\qquad \rho \equiv \sqrt{%
\frac{\dot{x}^{k}\dot{x}_{k}}{1+\lambda ^{2}\left( \kappa _{l}\kappa
^{l}+\partial _{l}\kappa ^{l}\right) }}\frac{\partial \left( \tau _{0},\tau
_{1},\tau ,\tau _{3}\right) }{\partial \left( x^{0},x^{1},x^{2},x^{3}\right)
}  \label{g1.24}
\end{equation}

The dynamic equation contains a collective variable $\rho ,$ describing
distribution of elements in the statistical ensemble, associated with the
dynamic system ${\cal S}_{{\rm KG}}$. The field of stochastic velocities $%
{\bf u}_{{\rm st}}$ is also connected with similar collective variable. One
can ignore this collective variable in the nonrelativistic case (\ref{g1.19}%
), because the cloud of pairs is coupled rigidly with the particle. Besides,
the variable $\rho ,$ defined by (\ref{g1.18a}), contains only transversal
derivatives, because in nonrelativistic case all space derivatives are
transversal. But one cannot ignore the collective variable $\rho $ in the
relativistic case (\ref{g1.23}), because the cloud of pairs can exist
separately from the particle. The pair production is a corollary of
independent existence of the pair cloud (but not of the quantization
procedure). It can be shown also mathematically \cite{R98}. Collective
(statistical) character of the $\kappa $ field, responsible for pair
production, does not permit one to separate one classical dynamic system
from the statistical ensemble, as it is possible in the nonrelativistic
case, when the pair production is neglected.

The Dirac particle is the dynamic system ${\cal S}_{{\rm D}}$, described by
the Dirac equation. The Dirac dynamic system ${\cal S}_{{\rm D}}$ was
investigated by many researchers. There is no possibility to list all them,
and we mention only some of them. First, this is transformation of the Dirac
equation on the base of quantum mechanics \cite{D58,FW50}. The complicated
structure of Dirac particle was discovered by Schr\"{o}dinger \cite{S930},
who interpreted it as some complicated quantum motion (zitterbewegung).
Investigation of this quantum motion and different models of Dirac particle
can be found in \cite{B84,BB81,A81,H90,RV93} and references therein. Our
investigation differs in absence of any suppositions on the Dirac particle
model and in a use of only dynamic methods.

The goal of investigation is a construction of dynamic system ${\cal S}_{%
{\rm dc}}$, whose dynamic equations form a system of ordinary differential
equations. Further the dynamic system ${\cal S}_{{\rm dc}}$, will be
referred to as classical Dirac particle. It has finite number of degrees of
freedom, and is simpler for investigation, than ${\cal S}_{{\rm D}}$.
D-disquantization of ${\cal S}_{{\rm D}}$ determines the classical dynamic
system ${\cal S}_{{\rm dc}}$ uniquely.

The first step of the investigation is transformation of WF-description into
HDP-description (the second and the third sections). In the
HDP-representation one realizes D-disquantization, transforming ${\cal S}_{%
{\rm D}}$ into ${\cal S}_{{\rm dc}}.$ It is the second step, described in
the fourth section. Writing actions for ${\cal S}_{{\rm D}}$ and ${\cal S}_{%
{\rm dc}}$ in the relativistically covariant form in the HDP-representation,
one discovers that both actions contain constant timelike 4-vector $f^{k}$,
describing splitting of space-time into space and time. It means that both
dynamic systems ${\cal S}_{{\rm D}}$ and ${\cal S}_{{\rm dc}}$ are
nonrelativistic. This unexpected difficulty is discussed in the fifth
section. One succeeded to overcome this obstacle, slightly modifying the
dynamic system ${\cal S}_{{\rm dc}}$. The unit timelike 4-vector $f^{k}$ is
identified with the constant 4-velocity of ${\cal S}_{{\rm dc}}$ (sixth
section). After such a modification the dynamic system ${\cal S}_{{\rm dc}}$
turns to relativistic dynamic system ${\cal S}_{{\rm dcr}}.$ In the seventh
section one solves dynamic equations for ${\cal S}_{{\rm dcr}}$. In the
eighth section one considers relativistic rotator ${\cal S}_{{\rm rr}}$ and
investigates its property. Comparing ${\cal S}_{{\rm dcr}}$ and ${\cal S}_{%
{\rm rr}}$, one discovers that ${\cal S}_{{\rm dcr}}$ is a relativistic
rotator and determines the rigidity function of ${\cal S}_{{\rm dcr}}$.
Calculations, connected with transformation from WF-representation to
HDP-representation, are rather bulky. They are presented in mathematical
appendices in detail. Unfortunately, detailed presentation of these
calculations is necessary, because their result (nonrelativistic character
of ${\cal S}_{{\rm dc}}$) seems to be doubtful for most of readers.

\section{Transformation of variables}

Considering dynamic system ${\cal S}_{{\rm D}}$, we set for simplicity that
the speed of the light $c=1$. ${\cal S}_{{\rm D}}$ is described in
WF-representation by the action
\begin{equation}
{\cal S}_{{\rm D}}:\qquad {\cal A}_{{\rm D}}[\bar{\psi},\psi ]=\int (-m\bar{%
\psi}\psi +{\frac{i}{2}}\hbar \bar{\psi}\gamma ^{l}\partial _{l}\psi -{\frac{%
i}{2}}\hbar \partial _{l}\bar{\psi}\gamma ^{l}\psi )d^{4}x  \label{f1.0}
\end{equation}
Here $\psi $ is four-component complex wave function, $\bar{\psi}=\psi
^{\ast }\gamma ^{0}$ is conjugate wave function, and $\psi ^{\ast }$ is the
Hermitian conjugate one. $\gamma ^{i}$, $i=0,1,2,3$ are $4\times 4$ complex
constant matrices, satisfying the relation
\begin{equation}
\gamma ^{l}\gamma ^{k}+\gamma ^{k}\gamma ^{l}=2g^{kl}I,\qquad k,l=0,1,2,3.
\label{f1.1}
\end{equation}
where $I$ is unit $4\times 4$ matrix. The action (\ref{f1.0}) generates
dynamic equation for the dynamic system ${\cal S}_{{\rm D}}$, known as Dirac
equation
\begin{equation}
i\hbar \gamma ^{l}\partial _{l}\psi -m\psi =0  \label{f1.2}
\end{equation}
and expressions for physical quantities: the 4-flux $j^{k}$ of particles and
energy-momentum tensor $T_{l}^{k}$%
\begin{equation}
j^{k}=\bar{\psi}\gamma ^{k}\psi ,\qquad T_{l}^{k}=\frac{i}{2}\left( \bar{\psi%
}\gamma ^{k}\partial _{l}\psi -\partial _{l}\bar{\psi}\cdot \gamma ^{k}\psi
\right)  \label{f1.3}
\end{equation}

The state of dynamic system ${\cal S}_{{\rm D}}$ is described by eight real
dependent variables (eight real components of four-component complex wave
function $\psi $). Using designations
\begin{equation}
\gamma _{5}=\gamma ^{0123}\equiv \gamma ^{0}\gamma ^{1}\gamma ^{2}\gamma
^{3},  \label{f1.9}
\end{equation}
\begin{equation}
{\bf \sigma }=\{\sigma _{1},\sigma _{2},\sigma _{3},\}=\{-i\gamma ^{2}\gamma
^{3},-i\gamma ^{3}\gamma ^{1},-i\gamma ^{1}\gamma ^{2}\}  \label{f1.10}
\end{equation}
let us make the change of variables
\begin{equation}
\psi =Ae^{i\varphi +{\frac{1}{2}}\gamma _{5}\kappa }e^{-{\frac{i}{2}}\gamma
_{5}\bsigma\bmeta}e^{{\frac{i\pi }{2}}\bsigma {\bf n}}\Pi  \label{f1.11}
\end{equation}
\begin{equation}
\psi ^{\ast }=A\Pi e^{-{\frac{i\pi }{2}}\bsigma {\bf n}}e^{-{\frac{i}{2}}%
\gamma _{5}\bsigma\bmeta}e^{-i\varphi -{\frac{1}{2}}\gamma _{5}\kappa }
\label{f1.12}
\end{equation}
where (*) means the Hermitian conjugation, and
\begin{equation}
\Pi ={\frac{1}{4}}(1+\gamma ^{0})(1+{\bf z\sigma }),\qquad {\bf z}%
=\{z^{\alpha }\}=\hbox{const},\qquad \alpha =1,2,3;\qquad {\bf z}^{2}=1
\label{f1.13}
\end{equation}
is a zero divisor. The quantities $A$, $\kappa $, $\varphi $, ${\bf \eta }%
=\{\eta ^{\alpha }\}$, ${\bf n}=\{n^{\alpha }\}$, $\alpha =1,2,3,\;$ ${\bf n}%
^{2}=1$ are eight real parameters, determining the wave function $\psi .$
These parameters may be considered as new dependent variables, describing
the state of dynamic system ${\cal S}_{{\rm D}}$. The quantity $\varphi $ is
a scalar, and $\kappa $ is a pseudoscalar. Six remaining variables $A,$ $%
{\bf \eta }=\{\eta ^{\alpha }\}$, ${\bf n}=\{n^{\alpha }\}$, $\alpha
=1,2,3,\;$ ${\bf n}^{2}=1$ can be expressed through the flux 4-vector $j^{l}=%
\bar{\psi}\gamma ^{l}\psi $ and spin 4-pseudovector
\begin{equation}
S^{l}=i\bar{\psi}\gamma _{5}\gamma ^{l}\psi ,\qquad l=0,1,2,3  \label{f1.13a}
\end{equation}
Because of two identities
\begin{equation}
S^{l}S_{l}\equiv -j^{l}j_{l},\qquad j^{l}S_{l}\equiv 0.  \label{f1.14}
\end{equation}
there is only six independent components among eight components of
quantities $j^{l},$ and $S^{l}.$ Let us make a change of variables in the
action (\ref{f1.0}), using substitution (\ref{f1.11}) -- (\ref{f1.13}).

Matrices $\gamma _{5}$, ${\bf \sigma }=\{\sigma _{\alpha }\},$ $\alpha
=1,2,3 $ are determined by relations (\ref{f1.9}), (\ref{f1.10}) have the
following properties
\begin{equation}
\gamma _{5}\gamma _{5}=-1,\qquad \gamma _{5}\sigma _{\alpha }=\sigma
_{\alpha }\gamma _{5},\qquad \gamma ^{0\alpha }\equiv \gamma ^{0}\gamma
^{\alpha }=-i\gamma _{5}\sigma _{\alpha },\qquad \alpha =1,2,3;  \label{b4.4}
\end{equation}
\begin{equation}
\left( \gamma ^{0}\right) ^{\ast }=\gamma ^{0},\qquad \left( \gamma ^{\alpha
}\right) ^{\ast }=-\gamma ^{\alpha },\qquad \gamma ^{0}{\bf \sigma }={\bf %
\sigma }\gamma ^{0},\qquad \gamma ^{0}\gamma _{5}=-\gamma _{5}\gamma ^{0}
\label{b4.5}
\end{equation}
According to equations (\ref{f1.1}), (\ref{f1.9}), (\ref{f1.10}) the
matrices ${\bf \sigma }=\{\sigma _{\alpha }\}$, $\alpha =1,2,3$ satisfy the
relation
\begin{equation}
\sigma _{\alpha }\sigma _{\beta }=\delta _{\alpha \beta }+i\varepsilon
_{\alpha \beta \gamma }\sigma _{\gamma },\qquad \alpha ,\beta =1,2,3
\label{a3.3}
\end{equation}
where $\varepsilon _{\alpha \beta \gamma }$ is the antisymmetric
pseudo-tensor of Levi-Chivita $(\varepsilon _{123}=1)$.

Using relations (\ref{b4.4}),(\ref{b4.5}), (\ref{a3.3}) and (\ref{f1.13}),
it is easy to verify that
\begin{eqnarray}
\Pi ^{2} &=&\Pi ,\qquad \gamma _{0}\Pi =\Pi ,\qquad {\bf z\sigma }\Pi =\Pi ,
\label{b4.5a} \\
\Pi \gamma _{5}\Pi &=&0,\qquad \Pi \sigma _{\alpha }\Pi =z^{\alpha }\Pi
,\qquad \alpha =1,2,3.  \label{a3.6}
\end{eqnarray}
Generally, the wave functions $\psi ,\psi ^{\ast }$ defined by (\ref{f1.12})
are $4\times 4$ complex matrices. In the proper representation, where $\Pi $
has the form
\begin{equation}
\Pi =\left( \matrix{1&0&0&0\cr 0&0&0&0\cr 0&0&0&0\cr 0&0&0&0}\right)
\label{a3.7}
\end{equation}
the $\psi ,\psi ^{\ast }$ have the form

\begin{equation}
\psi =\left( \matrix{\psi _1&0&0&0\cr \psi _2&0&0&0\cr \psi _3&0&0&0\cr \psi
_4&0&0&0}\right) ,\qquad \psi ^{\ast }=\left( \matrix{\psi _1^*&\psi _2^*&
\psi _3^*&\psi _4^*\cr 0&0&0&0\cr 0&0&0&0\cr 0&0&0&0}\right)  \label{a3.8}
\end{equation}
Their product $\psi ^{\ast }O\psi $ , where $O$ is an arbitrary $4\times 4$
matrix, has the form
\begin{equation}
\psi ^{\ast }O\psi =\left( \matrix{a&0&0&0\cr 0&0&0&0\cr 0&0&0&0\cr 0&0&0&0}%
\right) =a\Pi =\Pi a  \label{a3.9}
\end{equation}
where $a$ is a complex quantity. If $f$ is an analytical function having the
property $f(0)=0,$ then the function $f(\psi ^{\ast }O\psi )=f(a\Pi )$ of a $%
4\times 4$ matrix of the type (\ref{a3.9}) is a matrix $f(a)\Pi $ of the
same type. For this reason one will not distinguish between the complex
quantity $a$ and the complex $4\times 4$ matrix $a\Pi $. In the final
expressions of the type $a\Pi $ ($a$ is a complex quantity) the multiplier $%
\Pi $ will be omitted.

By means of relations (\ref{b4.4}) -- (\ref{a3.6}), one can reduce any
Clifford number $\Pi O\Pi $ to the form (\ref{a3.9}), without using any
concrete form of the $\gamma $-matrix representation. This property will be
used in our calculations. Calculating exponents of the type (\ref{f1.11}), (%
\ref{f1.12}), we shall use the following relations
\[
e^{-\frac{i\pi }{2}\bsigma {\bf n}}F\left( {\bf \sigma }\right) e^{\frac{%
i\pi }{2}\bsigma {\bf n}}=F\left( {\bf \Sigma }\right)
\]
where $F$ is arbitrary function and
\begin{equation}
{\bf \Sigma }=\{\Sigma _{1},\Sigma _{2},\Sigma _{3}\},\qquad \Sigma _{\alpha
}=e^{-\frac{i\pi }{2}\bsigma {\bf n}}\sigma _{\alpha }e^{\frac{i\pi }{2}%
\bsigma {\bf n}}\qquad \alpha =1,2,3;  \label{a3.12}
\end{equation}
satisfies the same commutation relations (\ref{a3.3}) as the Pauli matrices $%
{\bf \sigma }$.

For variables $\bar{\psi}\psi $, $j^{l}$, $S^{l}$, $l=0,1,2,3$ one has the
following expressions
\[
\bar{\psi}\psi =\psi ^{\ast }\gamma ^{0}\psi =A^{2}\Pi e^{\gamma _{5}\kappa
}\Pi =A^{2}\Pi \left( \cos \kappa +\gamma _{5}\sin \kappa \right) \Pi
=A^{2}\cos \kappa \Pi
\]
Taking into account the first relation (\ref{a3.6}), the term linear with
respect to $\gamma _{5}$ vanishes, and one obtains
\[
\bar{\psi}\psi =A^{2}\cos \kappa \Pi
\]
\begin{eqnarray}
j^{0}\Pi &=&\bar{\psi}\gamma ^{0}\psi =A^{2}\Pi e^{-\frac{i\pi }{2}\bsigma
{\bf n}}e^{-i\gamma _{5}\bsigma\bmeta}e^{\frac{i\pi }{2}\bsigma {\bf n}}\Pi
\nonumber \\
&=&A^{2}\Pi e^{-i\gamma _{5}{\bf \Sigma }\bmeta}\Pi =A^{2}\Pi \left( \cosh
\eta -\frac{i\gamma _{5}}{\eta }{\bf \Sigma \eta }\sinh \eta \right) \Pi
\nonumber \\
&=&A^{2}\cosh (\eta )\Pi  \label{f1.45}
\end{eqnarray}
where
\[
\eta =\sqrt{{\bf \eta }^{2}}=\sqrt{\eta ^{\alpha }\eta ^{\alpha }}
\]
Again in force of the first relation (\ref{a3.6}) we omit terms linear with
respect to $\gamma _{5}$.

In the same way one obtains
\begin{eqnarray*}
j^{\alpha }\Pi &=&\psi ^{\ast }\gamma ^{0\alpha }\psi \Pi =A^{2}\Pi e^{-{%
\frac{i}{2}}\gamma _{5}{\bf \Sigma }\bmeta}(-i\gamma _{5}\Sigma _{\alpha
})e^{-{\frac{i}{2}}\gamma _{5}{\bf \Sigma }\bmeta}\Pi = \\
&=&A^{2}\Pi (\cosh {\frac{\eta }{2}}-i\gamma _{5}{\bf \Sigma v}\sinh {\frac{%
\eta }{2}})(-i\gamma _{5}\Sigma _{\alpha })(\cosh {\frac{\eta }{2}}-i\gamma
_{5}{\bf \Sigma v}\sinh {\frac{\eta }{2}})\Pi = \\
&=&A^{2}\Pi (\cosh {\frac{\eta }{2}}\sinh {\frac{\eta }{2}}\left( \Sigma
_{\beta }\Sigma _{\alpha }+\Sigma _{\beta }\Sigma _{\alpha }\right) v^{\beta
}\Pi
\end{eqnarray*}
\begin{equation}
j^{\alpha }\Pi =A^{2}\sinh (\eta )v^{\alpha }\Pi ,\qquad \alpha =1,2,3
\label{a3.13}
\end{equation}
where
\begin{equation}
{\bf v}=\{v^{\alpha }\},\qquad v^{\alpha }=\eta ^{\alpha }/\eta ,\qquad
\alpha =1,2,3;\qquad {\bf v}^{2}=1.  \label{a3.14}
\end{equation}

Let us introduce designation ${\bf \xi }=\{\xi ^{\alpha }\}$, $\alpha =1,2,3$
for the expression
\begin{equation}
\xi ^{\alpha }\Pi =\Pi \Sigma _{\alpha }\Pi ,\qquad \alpha =1,2,3,\qquad
{\bf \xi }^{2}=\xi ^{\alpha }\xi ^{\alpha }=1  \label{a3.16}
\end{equation}
Then for the spin pseudovector $S^{l}$, defined by the relation (\ref{f1.13a}%
), one obtains

\begin{equation}
\begin{array}{lll}
S^{0}\Pi & = & \psi ^{\ast }(-i\gamma _{5})\psi =A^{2}\Pi (-i\gamma
_{5})e^{-i\gamma _{5}{\bf \Sigma }\bmeta}\Pi = \\
& = & A^{2}\Pi \sinh (\eta ){\bf \Sigma }{\bf v}\Pi =A^{2}\sinh (\eta ){\bf %
\xi v}\Pi ,
\end{array}
\label{a3.14a}
\end{equation}
\begin{eqnarray*}
S^{\alpha }\Pi &=&\psi ^{\ast }\gamma ^{0}i\gamma _{5}\gamma ^{\alpha }\psi
=\Pi \psi ^{\ast }\sigma _{\alpha }\psi \Pi =A^{2}\Pi e^{-{\frac{i}{2}}%
\gamma _{5}{\bf \Sigma }\bmeta}\Sigma _{\alpha }e^{-{\frac{i}{2}}\gamma _{5}%
{\bf \Sigma }\bmeta}\Pi = \\
&=&A^{2}\Pi (\cosh {\frac{\eta }{2}}-i\gamma _{5}{\bf \Sigma v}\sinh {\frac{%
\eta }{2}})\Sigma _{\alpha }(\cosh {\frac{\eta }{2}}-i\gamma _{5}{\bf \Sigma
v}\sinh {\frac{\eta }{2}})\Pi = \\
&=&A^{2}\Pi \left( \cosh ^{2}\frac{\eta }{2}\Sigma _{\alpha }+\sinh ^{2}{%
\frac{\eta }{2}}\left( \Sigma _{\beta }v^{\beta }\right) \Sigma _{\alpha
}\left( \Sigma _{\gamma }v^{\gamma }\right) \right) \Pi
\end{eqnarray*}
Now twice using relations (\ref{a3.3}) for Pauli matrices $\Sigma _{\alpha }$%
, one derives
\begin{eqnarray*}
S^{\alpha }\Pi &=&A^{2}\Pi \left( \cosh ^{2}\frac{\eta }{2}\Sigma _{\alpha
}+\sinh ^{2}{\frac{\eta }{2}}v^{\beta }v^{\gamma }\left( \delta _{\alpha
\beta }+i\varepsilon _{\beta \alpha \mu }\Sigma _{\mu }\right) \Sigma
_{\gamma }\right) \Pi = \\
&=&A^{2}\Pi \left( \cosh ^{2}\frac{\eta }{2}\Sigma _{\alpha }+\sinh ^{2}{%
\frac{\eta }{2}}\left( v^{\alpha }v^{\gamma }\Sigma _{\gamma }+i\varepsilon
_{\beta \alpha \mu }v^{\beta }v^{\gamma }\left( \delta _{\mu \gamma
}+i\varepsilon _{\mu \gamma \nu }\Sigma _{\nu }\right) \right) \right) \Pi \\
&=&A^{2}\Pi \left( \cosh ^{2}\frac{\eta }{2}\Sigma _{\alpha }+\sinh ^{2}{%
\frac{\eta }{2}}\left( v^{\alpha }v^{\gamma }\Sigma _{\gamma }-v^{\beta
}v^{\beta }\Sigma _{\alpha }+v^{\beta }v^{\alpha }\Sigma _{\beta }\right)
\right) \Pi
\end{eqnarray*}
\begin{equation}
S^{\alpha }\Pi =A^{2}[\xi ^{\alpha }+(\cosh \eta -1)v^{\alpha }({\bf v\xi }%
)]\Pi ,\qquad \alpha =1,2,3.  \label{a3.15}
\end{equation}

It follows from equations (\ref{f1.45}), (\ref{a3.13}), (\ref{a3.14})
\begin{equation}
j^{i}j_{i}\Pi =A^{4}\Pi ,\qquad A=(j^{l}j_{l})^{1/4}\equiv \rho ^{1/2}
\label{a3.17}
\end{equation}
According to the third equation (\ref{b4.5a}), (\ref{a3.12}) and (\ref{a3.16}%
) one obtains
\begin{eqnarray*}
\xi ^{\alpha }\Pi &=&\Pi \Sigma _{\alpha }\Pi =\Pi e^{-\frac{i\pi }{2}%
\bsigma {\bf n}}\sigma _{\alpha }e^{\frac{i\pi }{2}\bsigma {\bf n}}\Pi = \\
&=&\Pi \left( \cos \frac{\pi }{2}-i{\bf \sigma n}\sin \frac{\pi }{2}\right)
\sigma _{\alpha }\left( \cos \frac{\pi }{2}+i{\bf \sigma n}\sin \frac{\pi }{2%
}\right) \Pi = \\
&=&\Pi \left( {\bf \sigma n}\right) \sigma _{\alpha }\left( {\bf \sigma n}%
\right) \Pi =\Pi n^{\mu }n^{\nu }\sigma _{\mu }\sigma _{\alpha }\sigma _{\nu
}\Pi = \\
&=&\Pi \left( n^{\alpha }n^{\nu }\sigma _{\nu }+i\varepsilon _{\mu \alpha
\gamma }\sigma _{\gamma }\sigma _{\nu }n^{\mu }n^{\nu }\right) \Pi =\Pi
\left( n^{\alpha }n^{\nu }\sigma _{\nu }-\varepsilon _{\mu \alpha \gamma
}\varepsilon _{\gamma \nu \beta }\sigma _{\beta }n^{\mu }n^{\nu }\right) \Pi
= \\
&=&\Pi \left( n^{\alpha }n^{\nu }z_{\nu }-\varepsilon _{\mu \alpha \gamma
}\varepsilon _{\gamma \nu \beta }z^{\beta }n^{\mu }n^{\nu }\right) \Pi \\
&=&\left( n^{\alpha }\left( {\bf nz}\right) +\left( {\bf n}\times \left(
{\bf n}\times {\bf z}\right) \right) ^{\alpha }\right) \Pi ,\qquad \alpha
=1,2,3;
\end{eqnarray*}
Or

\begin{equation}
{\bf \xi }=2{\bf n}({\bf nz})-{\bf z}  \label{a3.21}
\end{equation}

\section{Transformation of the action}

The last two terms of the action (\ref{f1.0}) may be written in the form
\begin{eqnarray*}
{\frac{i}{2}}\hbar \bar{\psi}\gamma ^{l}\partial _{l}\psi +\text{h.c} &=&{%
\frac{i}{2}}\hbar \psi ^{\ast }\left( \partial _{0}-i\gamma _{5}{\bf \sigma
\nabla }\right) \psi +\text{h.c} \\
&=&{\frac{i}{2}}\hbar \psi ^{\ast }\left( \left( \partial _{0}-i\gamma _{5}%
{\bf \sigma \nabla }\right) \left( i\varphi +\frac{1}{2}\gamma _{5}\kappa
\right) \right) \psi +\text{h.c.} \\
&&+\frac{i}{2}\hbar A^{2}\Pi e^{-\frac{i\pi }{2}\bsigma {\bf n}}e^{-\frac{i}{%
2}\gamma _{5}\bsigma\bmeta}(\partial _{0}-i\gamma _{5}{\bf \sigma }\nabla
)(e^{-\frac{i}{2}\gamma _{5}\bsigma\bmeta}e^{\frac{i\pi }{2}\bsigma {\bf n}%
})\Pi +\text{h.c}
\end{eqnarray*}
where ''h.c.'' means the term obtained from the previous one by the
Hermitian conjugation. Calculation of this expression gives the following
result (see details of calculation in Appendix A). Let us set 
\begin{equation}
{\frac{i}{2}}\hbar \bar{\psi}\gamma ^{l}\partial _{l}\psi +\text{h.c}%
=F_{1}+F_{2}+F_{3}+F_{4}  \label{a4.1}
\end{equation}
where 
\begin{equation}
F_{1}+F_{2}=-j^{l}\partial _{l}\varphi \Pi -{\frac{1}{2}}\hbar S^{l}\partial
_{l}\kappa \Pi  \label{b5.5a}
\end{equation}
\begin{equation}
F_{3}=-\frac{\hbar j^{l}}{2\left( 1+{\bf \xi z}\right) }\varepsilon _{\alpha
\beta \gamma }\xi ^{\alpha }\partial _{l}\xi ^{\beta }z^{\gamma }\Pi
\label{b5.10}
\end{equation}
\begin{equation}
F_{4}=-{\frac{1}{2}}\hbar A^{2}\varepsilon _{\alpha \beta \gamma }\left(
\partial _{\alpha }\eta v^{\beta }+\sinh \eta \partial _{\alpha }v^{\beta
}+2\sinh ^{2}({\frac{\eta }{2}})v^{\alpha }\partial _{0}v^{\beta }\right)
\xi ^{\gamma }\Pi  \label{a4.6}
\end{equation}
where $\varepsilon _{\alpha \beta \gamma }$ is 3-dimensional Levi-Chivita
pseudotensor.

We see that the expressions (\ref{b5.5}) for $F_{1}$ and $F_{2}$ as well as
the first term of the action (\ref{f1.0}) 
\begin{equation}
-m\bar{\psi}\psi =-m\Pi e^{\gamma _{5}\kappa }\Pi =-mA^{2}\cos \kappa \Pi =-m%
\sqrt{j^{l}j_{l}}\cos \kappa \Pi \equiv -m\rho \cos \kappa \Pi  \label{b5.11}
\end{equation}
have relativistically covariant form. The terms $F_{3}$ and $F_{4}$ have
non-covariant form, and we try to write them in a covariant form. All these
terms contain non-covariant three-dimensional Levi-Chivita pseudotensor $%
\varepsilon _{\alpha \beta \gamma }$. It can be considered as spatial
components of the 4-dimensional Levi-Chivita pseudotensor $\varepsilon
_{iklm}\;\;(\varepsilon _{0123}=1)$, convoluted with the constant timelike
unit vector $f^{l}=\{1,0,0,0\}$. Then only spatial components of $%
\varepsilon _{iklm}f^{m}$ do not vanish 
\begin{equation}
\varepsilon _{\alpha \beta \gamma }=-\varepsilon _{\alpha \beta \gamma
m}f^{m},\qquad \alpha ,\beta ,\gamma =1,2,3  \label{b5.12}
\end{equation}
and one may substitute relation (\ref{b5.12}) in expression (\ref{b5.10})
for $F_{3}$.

\begin{equation}
F_{3}=\frac{\hbar j^{l}}{2\left( 1+{\bf \xi z}\right) }\varepsilon
_{iklm}\xi ^{i}\partial _{l}\xi ^{k}z^{l}f^{m}\Pi  \label{b5.14}
\end{equation}
where $\xi ^{l},\;z^{l}$ are 4-pseudovectors, whose spatial components in
the considered coordinate system are ${\bf \xi }$ and ${\bf z,}$ and
temporal components are of no importance. To write the scalar product ${\bf %
\xi z}$ in a covariant form, one sets $z^{0}=0$ in the considered coordinate
system.

Let us introduce constant 4-vector $f^{l}$ and constant 4-pseudovector $%
z^{l} $ by means of relations 
\[
f^{i}=\{1,0,0,0\},\qquad z^{i}=\{0,z^{1},z^{2},z^{3}\} 
\]
Then ${\bf \xi z}=-\xi _{l}z^{l}$, and one can rewrite the relation (\ref
{b5.14}) in the covariant form 
\begin{equation}
F_{3}=\hbar j^{l}\varepsilon _{iklm}\frac{\xi ^{i}}{\sqrt{2\left( 1-\xi
^{s}z_{s}\right) }}\partial _{l}\frac{\xi ^{k}}{\sqrt{2\left( 1-\xi
^{s}z_{s}\right) }}z^{l}f^{m}\Pi  \label{b5.15}
\end{equation}
One may introduce factor $\left[ 2\left( 1-\xi ^{s}z_{s}\right) \right]
^{-1/2}$ under sign of derivative, because differentiation of it gives $0$
in virtue of the vanishing factor $\varepsilon _{iklm}\xi ^{i}\xi
^{k}z^{l}f^{m}=0$.

Finally, introducing unit 4-pseudovector 
\[
\nu ^{i}=\xi ^{i}-(\xi ^{s}f_{s})f^{i},\qquad i=0,1,2,3;\qquad \nu ^{i}\nu
_{i}=-1, 
\]
and 4-pseudovector 
\[
\mu ^{i}\equiv \frac{\nu ^{i}}{\sqrt{-(\nu ^{l}+z^{l})(\nu _{l}+z_{l})}}=%
\frac{\nu ^{i}}{\sqrt{2(1-\nu ^{l}z_{l})}}=\frac{\nu ^{i}}{\sqrt{2(1+{\bf %
\xi z})}}. 
\]
one can rewrite the expression for $F_{3}$ in the form 
\begin{equation}
F_{3}=\hbar j^{l}\varepsilon _{iklm}\mu ^{i}\partial _{l}\mu
^{k}z^{l}f^{m}\Pi  \label{b5.16}
\end{equation}

Transformation of the expression (\ref{a4.6}) for $F_{4}$ to covariant form
is rather complicated. As a result of this transformation the expression (%
\ref{a4.6}) takes the form (See proof of this fact in Appendix B) 
\begin{equation}
F_{4}=-\frac{\hbar }{2(\rho +f^{s}j_{s})}\varepsilon _{iklm}[\partial
^{k}(j^{i}+f^{i}\rho )](j^{l}+f^{l}\rho )[\xi ^{m}-(\xi ^{s}f_{s})f^{m}]
\label{a4.9}
\end{equation}

Let us introduce the unit timelike vector 
\[
q^{i}\equiv \frac{j^{i}+f^{i}\rho }{\sqrt{(j^{l}+f^{l}\rho )(j_{l}+f_{l}\rho
)}}=\frac{j^{i}+f^{i}\rho }{\sqrt{2\rho (\rho +j^{l}f_{l})}},\qquad
q_{s}q^{s}=1 
\]
Then the relation (\ref{a4.9}) can be written shortly in a covariant form

\begin{equation}
F_{4}=\hbar \rho \varepsilon _{iklm}q^{i}(\partial ^{k}q^{l})\nu ^{m}
\label{c4..11}
\end{equation}

Now one can write the action (\ref{f1.0}) in the covariant form 
\begin{equation}
{\cal S}_{{\rm D}}:\qquad {\cal A}_{D}[j,\varphi ,\kappa ,{\bf \xi }]=\int 
{\cal L}d^{4}x,\qquad {\cal L}={\cal L}_{cl}+{\cal L}_{q1}+{\cal L}_{q2}
\label{c4.15}
\end{equation}
\begin{equation}
{\cal L}_{cl}=-m\rho -\hbar j^{i}\partial _{i}\varphi +\hbar
j^{s}\varepsilon _{iklm}\mu ^{i}\partial _{s}\mu ^{k}z^{l}f^{m},\qquad \rho
\equiv \sqrt{j^{l}j_{l}}  \label{c4.16}
\end{equation}
\begin{equation}
{\cal L}_{q1}=2m\rho \sin ^{2}({\frac{\kappa }{2}})-{\frac{\hbar }{2}}%
S^{l}\partial _{l}\kappa ,  \label{c4.17}
\end{equation}
\begin{equation}
{\cal L}_{q2}=-\hbar \rho \varepsilon _{iklm}q^{i}(\partial ^{k}q^{l})\nu
^{m}  \label{c4.18}
\end{equation}
Lagrangian is a function of 4-vector $j^{l}$, scalar $\varphi $,
pseudoscalar $\kappa $, and unit 3-pseudovector ${\bf \xi }$, which is
connected with the spin 4-pseudovector $S^{l}$ by means of the relation 
\begin{equation}
\xi ^{\alpha }=\rho ^{-1}\left[ S^{\alpha }-\frac{j^{\alpha }S^{0}}{%
(j^{0}+\rho )}\right] ,\qquad \alpha =1,2,3;\qquad \rho \equiv \sqrt{%
j^{l}j_{l}}  \label{f1.15}
\end{equation}
\begin{equation}
S^{0}={\bf j\xi },\qquad S^{\alpha }=\rho \xi ^{\alpha }+\frac{({\bf j\xi }%
)j^{\alpha }}{\rho +j^{0}},\qquad \alpha =1,2,3  \label{f1.16}
\end{equation}

\section{Dynamic disquantization and dynamic quantization}

Let ${\cal S}_{{\rm st}}$ be some stochastic system, and stochasticity of $%
{\cal S}_{{\rm st}}$ be a result of action of some stochastic agent on
deterministic dynamic system ${\cal S}_{{\rm d}}$. Let ${\cal E}\left[ {\cal %
S}_{{\rm st}}\right] $ and ${\cal E}\left[ {\cal S}_{{\rm d}}\right] $ be
statistical ensembles respectively of stochastic systems ${\cal S}_{{\rm st}%
} $ and deterministic dynamic systems ${\cal S}_{{\rm d}}$. Any statistical
ensemble is a distributed system, and all physical quantities $u=u\left( 
{\bf x}\right) $ are, in general, some functions of coordinates ${\bf x}$ of
the configuration space, where ${\cal E}\left[ {\cal S}_{{\rm st}}\right] $
is described. If the state of the statistical ensemble is such one, that the
quantities $u$ do not depend on ${\bf x}$, such a state will be referred to
as uniform state. If the quantities $u$ depend on ${\bf x}$ very slightly,
such a state will be referred to as quasi-uniform.

Dynamic equations for the statistical ensemble ${\cal E}\left[ {\cal S}_{%
{\rm st}}\right] $ describe the regular component of motion of stochastic
system ${\cal S}_{{\rm st}}$, constituting the statistical ensemble.
Stochastic component of motion of ${\cal S}_{{\rm st}}$ influences the
regular component of motion, provided the the ensemble state is not uniform,
and there are gradients of physical quantities $u$. Dynamic equations for
ensembles ${\cal E}\left[ {\cal S}_{{\rm st}}\right] $ and ${\cal E}\left[ 
{\cal S}_{{\rm d}}\right] $ coincide, if their state is uniform and
gradients of $u$ in dynamic equations for ${\cal E}\left[ {\cal S}_{{\rm st}}%
\right] $ vaniish. For instance, dynamic equations for the statistical
ensemble of Brownian particles have the form 
\[
\frac{\partial w}{\partial t}={\bf \nabla }\left( D{\bf \nabla }w-{\bf u}%
w\right) ,\qquad {\bf v}={\bf u}-D{\bf \nabla }\ln w 
\]
where distribution $w=w\left( {\bf x}\right) $ describes the state of the
ensemble. The velocity ${\bf u}={\bf u}\left( {\bf x}\right) $ describes
motion of the viscous medium, where the Brownian particles move. If the
state $w$ is uniform, and $w=$const, then ${\bf v}={\bf u},$ and the
velocity ${\bf v}$ of regular motion of stochastic Brownian particles
coincide with the velocity ${\bf u}$ of free particles, moving with a
friction in the viscous medium.

In the case of non-relativistic quantum system ${\cal S}_{{\rm S}}$,
described by the Schr\"{o}dinger equation, the system ${\cal S}_{{\rm S}}$
may be considered to be a dynamic system of the type ${\cal E}\left[ {\cal S}%
_{{\rm st}}\right] $. The state of ${\cal S}_{{\rm S}}$ is quasi-uniform, if
all physical quantities change slightly at the de Broglie wavelength. In
this case there exists such a coordinate system and such a state of ${\cal S}%
_{{\rm S}}$, where 
\begin{equation}
\lambda _{{\rm B}}|\frac{\partial u}{\partial x^{\alpha }}|\ll |u|,\qquad
\alpha =1,2,3,\qquad \lambda _{{\rm B}}=\frac{\hbar }{mv}  \label{c4.1}
\end{equation}
and the quantum dynamic system may be considered to be a classical one, in
the sense that dynamic equations for ${\cal E}\left[ {\cal S}_{{\rm st}}%
\right] $ coicide with those for ${\cal E}\left[ {\cal S}_{{\rm d}}\right] $%
. Here $u$ is any physical quantity and $\lambda _{{\rm B}}$ is the de
Broglie wavelength. In the relativistic case one cannot be sure, in general,
that there are such a coordinate system and such a state of the dynamic
system, where the condition (\ref{c4.1}) is fulfilled.

Let us introduce a local version of the condition (\ref{c4.1}), which is
written in the form 
\begin{equation}
\frac{\hbar }{mc}\sqrt{\frac{\left( j_{k}\partial _{\perp }^{k}u\right) ^{2}%
}{j^{s}j_{s}}}=\frac{\hbar }{mc}\sqrt{\partial _{k}u\cdot \partial ^{k}u-%
\frac{\left( j^{k}\partial _{k}u\right) ^{2}}{j^{s}j_{s}}}\ll |u|
\label{c4.2}
\end{equation}
where $c$ is the speed of the light, $j^{k}$ is the 4-vector of the particle
flux, and $\partial _{\perp }^{k}$ is transversal component of derivative,
defined by the relation (\ref{f1.4}). The condition (\ref{c4.2}) is somewhat
slighter, than the condition (\ref{c4.1}), because the Compton wavelength $%
\hbar /mc\leq \lambda _{{\rm B}}$. At the same time the condition (\ref{c4.2}%
) is relativistically covariant and local.

Let us introduce the procedure of dynamic disquantization
(D-disquantization). It is a special relativistically covariant procedure
which realizes the constraint (\ref{c4.2}). Any derivative $\partial ^{k}$
is separated into longitudinal component $\partial _{||}^{k}$ and
transversal one $\partial _{\perp }^{k},$ where $\partial _{||}^{k}$ and $%
\partial _{\perp }^{k}$ are defined by the relation (\ref{f1.4}). The
transversal component is neglected, and one obtains the dynamic system in
the quasi-uniform state.

For the quasi-uniform state, when one can neglect transversal derivatives,
the action (\ref{c4.15})--(\ref{c4.17}) takes the form 
\begin{eqnarray}
{\cal A}_{{\rm Dqu}}[j,\varphi ,\kappa ,{\bf \xi }] &=&\int \{-m\rho \cos
\kappa -\hbar j^{i}\left( \partial _{i}\varphi -\frac{\varepsilon _{jklm}\xi
^{j}\partial _{i}\xi ^{k}z^{l}f^{m}}{2\left( 1+{\bf \xi z}\right) }\right) 
\nonumber \\
&&-\frac{\hbar j^{i}}{{2\rho (}\rho +j^{j}f_{j}{)}}\varepsilon
_{klsm}j^{k}\partial _{i}j^{l}f^{s}\xi ^{m}\}d^{4}x  \label{a5.9}
\end{eqnarray}
Note that the second term $-\frac{\hbar }{2}S^{l}\partial _{l}\kappa $ in
the relation (\ref{c4.17}) is neglected, because 4-pseudovector $S^{k}$ is
orthogonal to 4-vector $j^{k}$, and the derivative is transversal.

Although the action (\ref{a5.9}) contains a non-classical variable $\kappa $%
, but in fact this variable is a constant. Indeed, a variation with respect
to $\kappa $ leads to the dynamic equation 
\begin{equation}
\frac{\delta {\cal A}_{Dqu}}{\delta \kappa }=m\rho \sin \kappa =0
\label{a5.10}
\end{equation}
which has solutions 
\begin{equation}
\kappa =n\pi ,\qquad n=\hbox{integer\qquad }  \label{a5.11}
\end{equation}
Thus, the effective mass $m_{{\rm eff}}=m\cos \kappa $ has two values 
\begin{equation}
m_{{\rm eff}}=m\cos \kappa =\pm m  \label{a5.12}
\end{equation}
The value $m_{{\rm eff}}=m>0$ $(\kappa ={\frac{\pi }{2}+2}n\pi )$
corresponds to a minimum of the action (\ref{a5.9}), whereas the value $m_{%
{\rm eff}}=-m<0$ corresponds to a maximum. Apparently, $m_{{\rm eff}}>0$
corresponds to a stable ensemble state, and $m_{{\rm eff}}<0$ does to
unstable state.

Eliminating $\kappa $ by means of the substitution $k={\frac{\pi }{2}+2}n\pi 
$ in (\ref{a5.9}), one obtains the action 
\[
{\cal A}_{{\rm Dqu}}[j,\varphi ,{\bf \xi }]=\int \{-m\rho -\hbar j^{i}\left(
\partial _{i}\varphi -\frac{\varepsilon _{jklm}\xi ^{j}\partial _{i}\xi
^{k}z^{l}f^{m}}{2\left( 1+{\bf \xi z}\right) }\right) 
\]
\begin{equation}
-\frac{\hbar j^{i}}{{2\rho (}\rho +j^{j}f_{j}{)}}\varepsilon
_{klsm}j^{k}\partial _{i}j^{l}f^{s}\xi ^{m}\}d^{4}x  \label{a5.13}
\end{equation}

Let us introduce Lagrangian coordinates $\tau =\{\tau _{0},{\bf \tau }%
\}=\{\tau _{i}\left( x\right) \}$, $i=0,1,2,3$ as functions of coordinates $%
x $ in such a way that only coordinate $\tau _{0}$ changes along the
direction $j^{l},$ i.e. 
\begin{equation}
j^{k}\partial _{k}\tau _{\mu }=0,\qquad \mu =1,2,3  \label{b3.1}
\end{equation}
Considering coordinates $x$ to be a functions of $\tau =\left\{ \tau _{0},%
{\bf \tau }\right\} $, one has the following identities 
\begin{equation}
\frac{\partial D}{\partial \tau _{0,k}}\tau _{i,k}\equiv \delta
_{i}^{0}D,\qquad i=0,1,2,3\qquad \tau _{i,k}\equiv \partial _{k}\tau
_{i},\qquad i,k=0,1,2,3  \label{b3.2}
\end{equation}
where 
\begin{equation}
D\equiv \frac{\partial (\tau _{0},\tau _{1},\tau _{2},\tau _{3})}{\partial
(x^{0},x^{1},x^{2},x^{3})},\qquad \frac{\partial D}{\partial \tau _{0,i}}%
\equiv \frac{\partial (x^{i},\tau _{1},\tau _{2},\tau _{3})}{\partial
(x^{0},x^{1},x^{2},x^{3})}.  \label{a5.14}
\end{equation}

Comparing (\ref{b3.1}) with (\ref{b3.2}), one concludes that it is possible
to set 
\begin{equation}
j^{i}=\frac{\partial D}{\partial \tau _{0,i}}\equiv \frac{\partial
(x^{i},\tau _{1},\tau _{2},\tau _{3})}{\partial (x^{0},x^{1},x^{2},x^{3})}%
,\qquad i=0,1,2,3  \label{b3.3}
\end{equation}
because the dynamic equation 
\begin{equation}
\frac{\delta {\cal A}_{{\rm Dqu}}}{\delta \varphi }=\hbar \partial
_{l}j^{l}=0  \label{b3.4}
\end{equation}
is satisfied by the relation (\ref{b3.3}) identically in force of identity 
\[
\partial _{i}\frac{\partial D}{\partial \tau _{k,i}}\equiv 0,\qquad
k=0,1,2,3. 
\]

Let us take into account that for any variable $u$%
\begin{equation}
D^{-1}j^{i}\partial _{i}u=D^{-1}\frac{\partial D}{\partial \tau _{0,i}}%
\partial _{i}u=\frac{\partial (u,\tau _{1},\tau _{2},\tau _{3})}{\partial
(\tau _{0},\tau _{1},\tau _{2},\tau _{3})}=\frac{du}{d\tau _{0}}
\label{b3.6}
\end{equation}
and in particular, 
\begin{equation}
D^{-1}j^{i}=D^{-1}\frac{\partial D}{\partial \tau _{0,i}}\equiv \frac{%
\partial (x^{i},\tau _{1},\tau _{2},\tau _{3})}{\partial (\tau _{0},\tau
_{1},\tau _{2},\tau _{3})}=\frac{dx^{i}}{d\tau _{0}}\equiv \dot{x}%
^{i},\qquad i=0,1,2,3  \label{b3.7}
\end{equation}
Besides 
\begin{equation}
d^{4}x=D^{-1}d^{4}\tau =D^{-1}d\tau _{0}d{\bf \tau }  \label{a5.16}
\end{equation}
\begin{equation}
j^{i}\partial _{i}\varphi =\frac{\partial (\varphi ,\tau _{1},\tau _{2},\tau
_{3})}{\partial (x^{0},x^{1},x^{2},x^{3})}  \label{a5.17}
\end{equation}

The action (\ref{a5.13}) can be rewritten in the Lagrangian coordinates $%
\tau $ in the form 
\begin{equation}
{\cal A}_{{\rm Dqu}}[x,{\bf \xi }]=\int \{-m\sqrt{\dot{x}^{i}\dot{x}_{i}}%
+\hbar {\frac{(\dot{{\bf \xi }}\times {\bf \xi }){\bf z}}{2(1+{\bf \xi z})}}%
+\hbar \frac{(\dot{{\bf x}}\times \ddot{{\bf x}}){\bf \xi }}{2\sqrt{\dot{x}%
^{s}\dot{x}_{s}}(\sqrt{\dot{x}^{s}\dot{x}_{s}}+\dot{x}^{0})}\}d^{4}\tau
\label{a5.18}
\end{equation}
where the dot means the total derivative $\dot{x}^{s}\equiv dx^{s}/d\tau
_{0} $.\ $x=\left\{ x^{0},{\bf x}\right\} =\{x^{i}\}$, $\;i=0,1,2,3$, ${\bf %
\xi }=\{\xi ^{\alpha }\}$, $\alpha =1,2,3$ are considered to be functions of
the Lagrangian coordinates $\tau _{0}$, ${\bf \tau }=\{\tau _{1},\tau
_{2},\tau _{3}\}$. $\;{\bf z}$ is the constant unit 3-vector (\ref{f1.13}).
The term $j^{i}\partial _{i}\varphi $ is omitted, because it reduces to a
Jacobian (\ref{a5.17}), which does not contribute to dynamic equations. In
fact, variables $x$ depend on ${\bf \tau }$ as on parameters, because the
action (\ref{a5.18}) does not contain derivatives with respect to $\tau
_{\alpha }$, \ $\alpha =1,2,3$. Lagrangian density of the action (\ref{a5.18}%
) does not contain independent variables $\tau $ explicitly. Hence, it may
be written in the form 
\begin{equation}
{\cal A}_{{\rm Dqu}}[x,{\bf \xi }]=\int {\cal A}_{{\rm dc}}[x,{\bf \xi }]d%
{\bf \tau ,\qquad d\tau }=d\tau _{1}d\tau _{2}d\tau _{3}  \label{b3.8}
\end{equation}
where 
\begin{equation}
{\cal A}_{{\rm dc}}[x,{\bf \xi }]=\int \{-m\sqrt{\dot{x}^{i}\dot{x}_{i}}%
+\hbar {\frac{(\dot{{\bf \xi }}\times {\bf \xi }){\bf z}}{2(1+{\bf \xi z})}}%
+\hbar \frac{(\dot{{\bf x}}\times \ddot{{\bf x}}){\bf \xi }}{2\sqrt{\dot{x}%
^{s}\dot{x}_{s}}(\sqrt{\dot{x}^{s}\dot{x}_{s}}+\dot{x}^{0})}\}d\tau _{0}
\label{b3.9}
\end{equation}

The action (\ref{b3.8}) is the action for the dynamic system ${\cal S}_{{\rm %
Dqu}},$ which is a set of similar independent dynamic systems ${\cal S}_{%
{\rm dc}}$. Such a dynamic system is called a statistical ensemble. Dynamic
systems ${\cal S}_{{\rm dc}}$ are elements (constituents) of the statistical
ensemble ${\cal E}_{{\rm Dqu}}$. Dynamic equations for each ${\cal S}_{{\rm %
dc}}$ form a system of ordinary differential equations. It may be
interpreted in the sense that the dynamic system ${\cal S}_{{\rm dc}}$ may
be considered to be a classical one, although its Lagrangian contains the
quantum constant $\hbar $.

The first term in the action (\ref{b3.9}) is relativistic. It describes a
motion of classical Dirac particle as a whole. The last two terms in the
action (\ref{b3.9}) are nonrelativistic. They describe some internal degrees
of freedom of the classical Dirac particle. This internal motion (classical
zitterbewegung) means that the classical Dirac particle has some internal
structure which is described by a method incompatible with relativity
principles. Maybe, the classical Dirac particle should be considered to be
consisting of several pointlike particles. At any rate the classical Dirac
particle is not a pointlike particle. It has a more complicated structure
which is described by the variable ${\bf \xi }$ and by the second order
derivative ${\bf \ddot{x}}$.

It is easy to see that the action (\ref{b3.9}) is invariant with respect to
transformation $\tau _{0}\rightarrow \tilde{\tau}_{0}=F\left( \tau
_{0}\right) $, where $F$ is an arbitrary monotone function. This
transformation admits one to choose the variable $t=x^{0}$ as a parameter $%
\tau _{0},$ or to choose the parameter $\tau _{0}$ in such a way that $\dot{x%
}^{l}\dot{x}_{l}=\dot{x}_{0}^{2}-{\bf \dot{x}}^{2}=1$. In the last case the
parameter $\tau _{0}$ is the proper time along the world line of classical
Dirac particle.

One can introduce the dynamic quantization (D-quantization) as a procedure
reciprocal to the D-disquantization. From mathematical viewpoint
D-quantization means introduction of transversal derivatives into the
action. Such an introduction of transversal derivatives means appearance in
the action of some additional terms, describing interaction between the
independent systems ${\cal S}_{{\rm dc}}$ of the statistical ensemble ${\cal %
E}_{{\rm Dqu}}$. There is a lot of different ways to introduce terms,
containing transversal derivatives. Each such introduction means
D-quantization of classical system. It may be considered to be an
alternative to canonical quantization, which is determined uniquely by the
Hamiltonian of the system and can be performed only for Hamiltonian dynamic
system. D-quantization is not unique. It is chosen on the base of some model
(physical) reasoning.

Application of D-quantization to the statistical ensemble ${\cal E}_{{\rm Dqu%
}}$ means that elements ${\cal S}_{{\rm dc}}$ of the statistical ensemble $%
{\cal E}_{{\rm Dqu}}$ becomes to interact between themselves through some
''quantum vector field'' $\kappa ^{i}$ and ''scalar quantum field'' $\kappa $%
. Of course, the dynamic quantization is not unique, because one can
introduce quantum fields $\kappa $ and $\kappa ^{i}$ in different ways. But
the quantum fields $\kappa ^{i}$ and $\kappa $ can be introduced in such a
way, that the dynamic system, formed by the set of interacting identical
dynamic systems ${\cal S}_{{\rm dc}}$, coincides with the dynamic system $%
{\cal S}_{{\rm D}}$. The D-quantization is a purely dynamic procedure which
is relativistically covariant. It admits one to reduce a statistical
ensemble ${\cal S}_{{\rm Dqu}}$ of classical dynamic systems to a quantum
system ${\cal S}_{{\rm D}}$ by means of dynamic methods, i.e. without a
reference to attributes of quantum mechanics.

Let us derive $\kappa $ and $\kappa ^{i}$ fields for the dynamic system $%
{\cal S}_{{\rm D}}$. If one adds the omitted terms 
\[
m\rho \left( 1-\cos \kappa \right) -\frac{\hbar }{2}S^{l}\partial _{l}\kappa
,\qquad \hbar \rho \varepsilon _{iklm}q^{i}\left( \partial _{\perp
}^{k}q^{l}\right) \nu ^{m} 
\]
to the Lagrangian density of (\ref{a5.13}), one returns to the action (\ref
{c4.15}) -- (\ref{c4.18}) for ${\cal S}_{{\rm D}}$. Let us make this. One
obtains the action (\ref{c4.15}) -- (\ref{c4.18}). Let us introduce new
variables 
\begin{equation}
\kappa ^{i}=q^{i}=\frac{j^{i}+\rho f^{i}}{\sqrt{2\rho (\rho +j^{s}f_{s})}}=%
\frac{j^{i}+\rho f^{i}}{\sqrt{(j^{s}+\rho f^{s})(j_{s}+\rho f_{s})}},\qquad
i=0,1,2,3;  \label{a5.23}
\end{equation}
and introduce them in the Lagrangian density (\ref{c4.18}) by means of the
Lagrangian multipliers $\lambda _{i}$. Then ${\cal L}_{q2}$ is substituted
by 
\begin{equation}
{\cal L}_{q2}^{\prime }=-\hbar \rho \varepsilon _{iklm}(\partial _{\perp
}^{k}\kappa ^{l})\kappa ^{i}\nu ^{m}-{\frac{\hbar }{\rho }}j^{s}\varepsilon
_{iklm}q^{i}j^{k}\partial _{s}q^{l}\nu ^{m}+\lambda _{i}\left( q^{i}-\kappa
^{i}\right)  \label{a5.24}
\end{equation}
where $\partial _{\perp }^{k}$ is defined by (\ref{f1.4}) and 
\begin{equation}
\rho \equiv \sqrt{j^{l}j_{l}},\qquad \nu ^{m}=[\xi ^{m}-(\xi
^{s}f_{s})f^{m}],\qquad m=0,1,2,3  \label{a5.25}
\end{equation}

The derivative $\partial ^k$ in (\ref{a5.24}) is separated into longitudinal
component $\partial ^k_{||}$ and transversal one $\partial ^k_{\perp }$. The
quantities $q^i$ in the coefficient at $\partial ^k_{\perp }$ are replaced
by $\kappa ^i$, in the coefficient at $\partial ^k_{||}$ they are not
changed. According to (\ref{a5.23}) $\kappa ^i=q^i$, and such a replacement
does not change anything essentially.

Variation of the action with respect to $\kappa ^{i}$ leads to the dynamic
equations 
\begin{equation}
-\hbar \rho \varepsilon _{iklm}(\partial _{\perp }^{k}\kappa ^{l})\nu
^{m}-\hbar \partial _{\perp }^{\ast k}\left( \rho \varepsilon _{iklm}\kappa
^{l}\nu ^{m}\right) +\lambda _{i}=0  \label{c4.12}
\end{equation}
where the operator $\partial _{\perp }^{\ast k}$ is defined by the relation 
\begin{equation}
\partial _{\perp }^{\ast k}u=\partial ^{k}u-\partial _{s}({\frac{j^{k}j^{s}}{%
\rho ^{2}}}u)  \label{a5.31}
\end{equation}

Eliminating $\lambda _{i}$ from (\ref{a5.24}) by means of (\ref{c4.12}), one
obtains instead of ${\cal L}_{q2}^{\prime }$ 
\begin{equation}
{\cal L}_{q3}=-\hbar \rho \varepsilon _{iklm}(\kappa ^{l}\partial _{\perp
}^{k}\kappa ^{i}-q^{l}\partial _{\perp }^{k}\kappa ^{i}+\kappa ^{i}\partial
_{\perp }^{k}q^{l})\nu ^{m}-\frac{\hbar j^{i}}{{2\rho (}\rho +j^{j}f_{j}{)}}%
\varepsilon _{klsm}j^{k}\partial _{i}j^{l}f^{s}\xi ^{m}  \label{a5.28}
\end{equation}
Now the action has the form 
\begin{equation}
{\cal A}_{D}[j,\varphi ,\kappa ,\xi ,\kappa ^{i}]=\int ({\cal L}_{cl}+{\cal L%
}_{q1}+{\cal L}_{q3})d^{4}x  \label{a5.29}
\end{equation}
where the Lagrangian densities ${\cal L}_{cl}$, ${\cal L}_{q1}$, ${\cal L}%
_{q3}$ are determined by (\ref{c4.16}), (\ref{c4.17}), (\ref{a5.28})
respectively. Dynamic equations generated by action (\ref{a5.29}) are
equivalent to dynamic equations, generated by actions (\ref{f1.0}) and (\ref
{c4.15}) -- (\ref{c4.18}). The action (\ref{a5.29}) generates the dynamic
equation 
\begin{equation}
{\frac{\delta {\cal A}_{D}}{\delta \kappa ^{i}}}=-\hbar \varepsilon
_{iklm}\{\rho \nu ^{m}\partial _{\perp }^{k}(q^{l}-\kappa ^{l})+\partial
_{\perp }^{\ast k}[\rho \nu ^{m}(q^{l}-\kappa ^{l})]\}=0  \label{a5.30}
\end{equation}
where the operator $\partial _{\perp }^{\ast k}$ is defined by the relation (%
\ref{a5.31}). Resolving (\ref{a5.30}) with respect to $\kappa ^{i}$ and
substituting the $\kappa ^{i}$ into (\ref{a5.29}), one returns to the action
(\ref{c4.15}) -- (\ref{c4.18}). The fact that the solution (\ref{a5.23}) of (%
\ref{a5.30}) is not unique is of no importance, because (\ref{a5.28})
reduces to (\ref{c4.18}) by virtue of (\ref{a5.30}). Indeed, convoluting
equation.(\ref{a5.30}) with $\kappa ^{i}$ and using the obtained relation
for eliminating $\kappa ^{i}$ from (\ref{a5.24}), one obtains (\ref{c4.18}).

Let us note that $\kappa ^{l}$ are not rigorous dynamic variables, because
the dynamic equations (\ref{a5.30}) for $\kappa ^{l}$ contain derivatives
only along spacelike directions orthogonal to $j^{i}$. Rather the
introduction of $\kappa ^{i}$ is an invariant (with respect to a change of
variables) way of separating out the classical part of the action.

The variables $\kappa ,$ $\kappa ^{i}$, \ $i=0,1,2,3$ are special quantum
variables, which are responsible for quantum effects, described by the Dirac
equation. They are introduced in such a way that, when they vanish, all
terms, containing transversal component of derivative vanish also, and $%
{\cal S}_{{\rm D}}$ reduces to ${\cal S}_{{\rm Dqu}}$. Indeed, let us
suppress quantum variables $\kappa ,$ $\kappa ^{i}$, \ $i=0,1,2,3$ in the
action (\ref{a5.28}), (\ref{a5.29}). Then the action (\ref{a5.29}) reduces
to the action (\ref{a5.13}).

\section{Relativistic invariance}

It is a common practice to think that if dynamic equations of a system can
be written in the relativistically covariant form, such a possibility
provides automatically relativistic character of considered dynamic system,
described by these equations. In general, it is valid only in the case, when
dynamic equations do not contain absolute objects, or these absolute objects
has the Poincare group as a group of their symmetry \cite{A67}. The absolute
object is one or several quantities, which are the same for all states of
the dynamic system \cite{A67}. A given external field, or metric tensor
(when it is given, but not determined from the gravitational equations) are
examples of absolute objects. In the case of dynamic system ${\cal S}_{{\rm D%
}}$, described by the Dirac equation the Dirac $\gamma $-matrices are
absolute objects.

Anderson \cite{A67} investigated in details the role of absolute objects for
symmetry of dynamic systems. His conclusion is as follows. If a dynamic
system is described by dynamic equations, written in covariant form, the
symmetry group of dynamic system is determined by the symmetry group of
these absolute objects. Here we confirm this result in a simple example,
when dynamic equations of certainly nonrelativistic dynamic system are
written in a relativistically covariant form.

Let us consider a system of differential equations, consisting of the
Maxwell equations for the electromagnetic tensor $F^{ik}$ in some inertial
coordinates $x$

\begin{equation}
\partial _{k}F^{ik}\left( x\right) =4\pi J^{i},\qquad \varepsilon
_{iklm}g^{jm}\partial _{j}F^{kl}\left( x\right) =0,\qquad \partial
_{k}\equiv \frac{\partial }{\partial x^{k}}  \label{a6.1}
\end{equation}
and equations 
\begin{equation}
m\frac{d}{d\tau }[(l_{k}\dot{q}^{k})^{-1}\dot{q}^{i}-{\frac{1}{2}}%
g^{ik}l_{k}(l_{j}\dot{q}^{j})^{-2}\dot{q}^{s}g_{sl}\dot{q}^{l}%
]=eF^{il}(q)g_{lk}\dot{q}^{k};\qquad i=0,1,2,3  \label{a6.2}
\end{equation}
\begin{equation}
\dot{q}^{k}\equiv \frac{dq^{k}}{d\tau }
\end{equation}
where $q^{i}=q^{i}(\tau )$, $i=0,1,2,3$ describe coordinates of a pointlike
charged particle as functions of a parameter $\tau $, $l_{i}$ is a constant
timelike unit vector, 
\begin{equation}
g^{ik}l_{i}l_{k}=1;  \label{a6.3}
\end{equation}
and the speed of the light $c=1$.

This system of equations is relativistically covariant with respect to
quantities $q^{i}$, $F^{ik}$, $J^{i}$, $l_{i}$, $g_{ik}$, i.e. it does not
change its form at any Lorentz transformation, which is accompanied by
corresponding transformation of quantities $q^{i}$, $F^{ik}$, $J^{i}$, $%
l_{i} $, $g_{ik}$, where the quantities $q^{i}$, $J^{i}$, $l_{i}$ transform
as 4-vectors and the quantities $F^{ik}$, $g_{ik}$ as 4-tensors

The reference to the quantities $q^{i}$, $F^{ik}$, $J^{i}$, $l_{i}$, $g_{ik}$
means that all these quantities are considered as formal dependent
variables, when one compares the form of dynamic equations written in two
different coordinate systems. For instance, if a reference to $J^{i}$ is
omitted in the formulation of the relativistic covariance, it means that
components of $J^{i}$ are considered as some functions of the coordinates $x$%
. If $J^{i}\neq 0$, then $J^{i}$ and $\tilde{J}^{i}$ in other coordinate
system are different functions of the arguments $x$ and $\tilde{x}$
respectively, and the first equation (\ref{a6.1}) has different form in
different coordinate systems. In other words, the dynamic equations (\ref
{a6.1})--(\ref{a6.2}) are not relativistically covariant with respect to
quantities $q^{i}$, $F^{ik}$, $l_{i}$, $g_{ik}$. Thus, for the relativistic
covariance it is important both the laws of transformation and how each of
quantities is considered as a formal variable, or as some function of
coordinates.

Following Anderson \cite{A67} we divide the quantities $q^{i}$, $F^{ik}$, $%
J^{i}$, $l_{i}$, $g_{ik}$ into two parts: dynamic objects (variables) $q^{i}$%
, $F^{ik}$ and absolute objects $J^{i}$, $l_{i}$, $g_{ik}$. By definition of
absolute objects they have the same value for all solutions of the dynamic
equations, whereas dynamic variables are different, in general, for
different solutions. If the dynamic equations are written in the
relativistically covariant form, their symmetry group (and a compatibility
with the relativity principles) is determined by the symmetry group of the
absolute objects $J^{i}$, $l_{i}$, $g_{ik}$.

Let for simplicity $J^{i}\equiv 0$. A symmetry group of the constant
timelike vector $l_{i}$ is a group of rotations in the $3$-plane orthogonal
to the vector $l_{i}$. The Lorentz group is a symmetry group of the metric
tensor $g_{ik}=$diag $\{1,-1,-1,-1\}$. Thus, the symmetry group of all
absolute objects $l_{i}$, $g_{ik}$, $J^{i}\equiv 0$ is a subgroup of the
Lorentz group (rotations in the $3$-plane orthogonal to $l_{i}$). As far as
the symmetry group is a subgroup of the Lorentz group and does not coincide
with it, the system of equations (\ref{a6.1})--(\ref{a6.2}) is
nonrelativistic (incompatible with the relativity principles).

Of course, the compatibility with the relativity principles does not depend
on the fact with respect to which quantities the relativistic covariance is
considered. For instance, let us consider a covariance of equations (\ref
{a6.1}), (\ref{a6.2}) with respect to quantities $q^{i}$, $F^{ik}$, $%
J^{i}\equiv 0$. It means that now $l_{i}$ are to be considered as functions
of $x$ (in the given case these functions are constants), because a
reference to $l_{i}$ as a formal variables is absent. After the
transformation to another coordinate system the equation (\ref{a6.2}) takes
the form 
\begin{equation}
m\frac{d}{d\tau }[(\tilde{l}_{k}\frac{d\tilde{q}}{d\tau }^{k})^{-1}\frac{d%
\tilde{q}^{i}}{d\tau }-\frac{1}{2}g^{ik}{\tilde{l}_{k}}(\tilde{l}_{j}\frac{d%
\tilde{q}^{j}}{d\tau })^{-2}\frac{d\tilde{q}}{d\tau }^{s}\frac{d\tilde{q}_{s}%
}{d\tau }]=e\tilde{F}^{il}(\tilde{q})g_{lk}\frac{d\tilde{q}^{k}}{d\tau }
\label{a6.10}
\end{equation}

Here $\tilde{l}_{i}$ are considered as functions of $\tilde{x}$. But $\tilde{%
l}_{i}$ are other functions of $\tilde{x}$, than $l_{i}$ of $x$ (other
numerical constants $\tilde{l}_{k}=l_{j}\partial x^{j}/\partial \tilde{x}%
^{k} $ instead of $l_{k}$), and equations (\ref{a6.2}) and (\ref{a6.10})
have different forms with respect to quantities $q^{i}$, $F^{ik}$, $%
J^{i}\equiv 0$. It means that (\ref{a6.2}) is not relativistically covariant
with respect to $q^{i}$, $F^{ik}$, $J^{i}\equiv 0$, although it is
relativistically covariant with respect to $q^{i}$, $F^{ik}$, $l_{i}$, $%
J^{i}\equiv 0$.

Setting $l_{i}=\{1,0,0,0\}$, $t=q^{0}(\tau )$ in (\ref{a6.2}), one obtains 
\begin{equation}
m\frac{d^{2}q}{dt^{2}}^{\alpha }=eF_{.0}^{\alpha }+eF_{.\beta }^{\alpha }%
\frac{dq}{dt}^{\beta },\qquad i=\alpha =1,2,3;  \label{a6.11}
\end{equation}
\[
\frac{m}{2}\frac{d}{dt}(\frac{dq}{dt}^{\alpha }\frac{dq}{dt}^{\alpha
})=eF_{.0}^{\alpha }\frac{dq}{dt}^{\alpha },\qquad i=0. 
\]
It is easy to see that this equation describes a nonrelativistic motion of a
charged particle in a given electromagnetic field $F^{ik}$. The fact that
the equations (\ref{a6.2}) or (\ref{a6.11}) are nonrelativistic is connected
with the space-time splitting into space and time that is characteristic for
Newtonian mechanics. This space-time splitting is described in different
ways in equations (\ref{a6.2}) and (\ref{a6.11}). It is described by the
constant timelike vector $l_{i}$ in (\ref{a6.2}). In the equation (\ref
{a6.11}) the space-time splitting is described by a special choice of the
coordinate system whose time axis is directed along the vector $l^{i}$.

This example shows that nonrelativistic equation (\ref{a6.11}) can be
written in a relativistically covariant form (\ref{a6.2}), provided one
introduces an absolute object $l_{i}$, describing space-time splitting.

Dirac matrices $\gamma ^{k}$ are absolute objects, as well as the metric
tensor $g^{kl}$, which may be considered as a derivative absolute object
determined by the relation (\ref{f1.1}).

There are two approaches to the Dirac equation. In the first approach \cite
{S30,S51} the wave function $\psi $ is considered to be a scalar function
defined on the field of Clifford numbers $\gamma ^{l}$, 
\begin{equation}
\psi =\psi (x,\gamma )\Gamma ,\qquad \overline{\psi }=\Gamma \overline{\psi }%
(x,\gamma ),  \label{b1.4}
\end{equation}
where $\Gamma $ is a constant nilpotent factor which has the property $%
\Gamma f(\gamma )\Gamma =a\Gamma $. Here $f(\gamma )$ is arbitrary function
of $\gamma ^{l}$ and $a$ is a complex number, depending on the form of the
function $f$. Within such an approach $\psi $, $\bar{\psi}$ transform as
scalars and $\gamma ^{l}$ transform as components of a 4-vector under the
Lorentz transformations. In this case the symmetry group of $\gamma ^{l}$ is
a subgroup of the Lorentz group, and ${\cal S}_{{\rm D}}$ is nonrelativistic
dynamic system. Then the matrix vector $\gamma ^{l}$ describes some
preferred direction in the space-time.

In the second (conventional) approach $\psi $ is considered to be a spinor,
and $\gamma ^{l},\quad l=0,1,2,3$ are scalars with respect to the
transformations of the Lorentz group. In this case the symmetry group of the
absolute objects $\gamma ^{l}$ is the Lorentz group, and dynamic system $%
{\cal S}_{{\rm D}}$ is considered to be a relativistic dynamic system.

Of course, the approaches leading to incompatible conclusions cannot be both
valid. At least, one of them is wrong. Analyzing the two approaches,
Sommerfeld \cite{S51} considered the first approach to be more reasonable.
In the second case the analysis is rather difficult due to non-standard
transformations of $\gamma ^{l}$ and $\psi $ under linear coordinate
transformations $T$. Indeed, the transformation $T$ for the vector $j^{l}=%
\bar{\psi}\gamma ^{l}\psi $ has the form 
\begin{equation}
\tilde{\overline{\psi }}\tilde{\gamma}^{l}\tilde{\psi}=\frac{\partial \tilde{%
x}^{l}}{\partial x^{s}}\bar{\psi}\gamma ^{s}\psi ,  \label{b1.5}
\end{equation}
where quantities marked by tilde mean quantities in the transformed
coordinate system. This transformation can carried out by two different ways 
\begin{equation}
1:\;\;\;\tilde{\psi}=\psi ,\qquad \tilde{\overline{\psi }}=\overline{\psi }%
,\qquad \tilde{\gamma}^{l}=\frac{\partial \tilde{x}^{l}}{\partial x^{s}}%
\gamma ^{s},\qquad l=0,1,2,3  \label{b1.6}
\end{equation}
\begin{equation}
2:\;\;\;\tilde{\gamma}^{l}=\gamma ^{l},\qquad l=0,1,2,3,\qquad \tilde{\psi}%
=S(\gamma ,T)\psi ,\qquad \tilde{\overline{\psi }}=\overline{\psi }%
S^{-1}(\gamma ,T),  \label{b1.7}
\end{equation}
\begin{equation}
S^{\ast }(\gamma ,T)\gamma ^{0}=\gamma ^{0}S^{-1}(\gamma ,T)  \label{b1.8}
\end{equation}
The relations (\ref{b1.6}) correspond to the first approach and the
relations (\ref{b1.7}) correspond to the second one. Both ways (\ref{b1.6})
and (\ref{b1.7}) lead to the same result, provided 
\begin{equation}
S^{-1}(\gamma ,T)\gamma ^{l}S(\gamma ,T)=\frac{\partial \tilde{x}^{l}}{%
\partial x^{s}}\gamma ^{s}  \label{b1.9}
\end{equation}
In particular, for infinitesimal Lorentz transformation $x^{i}\rightarrow
x^{i}+\delta \omega _{.k}^{i}x^{k}$ $S(\gamma ,T)$ has the form \cite{S61} 
\begin{equation}
S(\gamma ,T)=\exp \left( \frac{\delta \omega _{ik}}{8}\left( \gamma
^{i}\gamma ^{k}-\gamma ^{k}\gamma ^{i}\right) \right)  \label{b1.10}
\end{equation}
The second way (\ref{b1.7}) has two defects. First, the transformation law
of $\psi $ depends on $\gamma $, i.e. under linear transformation $T$ of
coordinates the components of $\psi $ transform through $\psi $ and $\gamma
^{l}$, but not only through $\psi $. Note that tensor components in a
coordinate system transform only through tensor components in other
coordinate system, and this transformation does not contain any absolute
objects. (for instance, the relation (\ref{b1.5})). Second, the relation (%
\ref{b1.9}) is compatible with (\ref{f1.1}) only under transformations $T$
between orthogonal coordinate systems, when components $g^{lk}=\{1,-1,-1,-1%
\} $ of the metric tensor are invariant. In other words, at the second
approach the relation (\ref{f1.1}) is not covariant, in general, with
respect to arbitrary linear transformations of coordinates. In this case one
cannot be sure that the symmetry group of the dynamic system coincides with
the symmetry group of absolute objects.

The fact that the symmetry group of dynamic system coincides with the
symmetry group of absolute objects was derived with the supposition, that
under the coordinate transformation any object transforms only via its
components. This condition is violated in the second case, and one cannot be
sure that the symmetry group of dynamic system coincides with that of
absolute objects.

After change of variables the action (\ref{f1.0}) transforms to the form (%
\ref{c4.15}) -- (\ref{c4.18}), the $\gamma $-matrices being eliminated. But
after reduction of the action to the relativistically covariant form two new
absolute objects appear: constant 4-vector $f^{i}$ and constant
4-pseudovector $z^{i}$. We shall see further that the 4-pseudovector $z^{i}$
is fictitious. But the action is really depends on 4-vector $f^{i}$, which
resembles the vector $l_{i}$ in the considered example (\ref{a6.2}). It
means that the dynamic system ${\cal S}_{{\rm D}}$ is nonrelativistic,
because it supposes an absolute separation of the space-time into the space
and the time.

The fact that the Dirac $\gamma $-matrices contain a constant timelike
vector seems to be rather unexpected. As far as calculations leading to this
result are rather bulky, they can cast some doubt upon their correctness. To
remove this doubt, we present detailed calculations in the mathematical
appendices. Besides a constant timelike vector appears to be ''hidden'' in
Dirac $\gamma $-matrices in the case of two-dimensional Dirac equation,
where corresponding calculations are essentially simpler \cite{R999}.

\section{Relativization of dynamic equations}

Let us return to the action (\ref{b3.9}), which is written in the covariant
form 
\begin{equation}
{\cal S}_{{\rm dc}}:{\rm {\cal A}}_{{\rm dc}}\left[ x,\xi \right] =\int \{-m%
\sqrt{\dot{x}^{i}\dot{x}_{i}}-\hbar \frac{\varepsilon _{iklm}\xi ^{i}\dot{\xi%
}^{k}f^{l}z^{m}}{2(1-\xi ^{s}z_{s})}-\frac{\hbar }{2}Q\varepsilon _{iklm}%
\dot{x}^{i}\ddot{x}^{k}f^{l}\xi ^{m}\}d\tau _{0}  \label{f5.1}
\end{equation}
where 
\begin{equation}
Q=Q\left( \dot{x}\right) =\frac{1}{\sqrt{\dot{x}^{s}\dot{x}_{s}}(\dot{x}%
^{l}f_{l}+\sqrt{\dot{x}^{l}\dot{x}_{l}})}  \label{f5.2}
\end{equation}
To relativize this action, it is sufficient to make the 4-vector $f^{i}$ to
be a dynamic variable instead of an absolute object. Let us identify $f^{i}$
with the constant 4-velocity vector 
\begin{equation}
u_{i}=-\frac{P_{i}}{M},\qquad M=\sqrt{P_{k}P^{k}}  \label{f5.3}
\end{equation}
where $P_{k}$ is the total momentum of the dynamic system ${\cal S}_{{\rm dc}%
}$, and $M$ is the total mass of ${\cal S}_{{\rm dc}}$. Let $L$ be
Lagrangian of the action (\ref{f5.1}). Then the momentum has the form 
\begin{eqnarray}
P_{i} &=&\frac{\partial L}{\partial \dot{x}^{i}}-\frac{d}{d\tau _{0}}\frac{%
\partial L}{\partial \ddot{x}^{i}}=-m\frac{\dot{x}_{i}}{\sqrt{\dot{x}^{s}%
\dot{x}_{s}}}-\frac{\hbar }{2}Q\varepsilon _{iklm}\ddot{x}^{k}f^{l}\xi ^{m} 
\nonumber \\
&&-\frac{\hbar }{2}\frac{d}{d\tau _{0}}\left( Q\varepsilon _{iklm}\dot{x}%
^{k}\xi ^{m}\right) f^{l}-\frac{\hbar }{2}\varepsilon _{sklm}\dot{x}^{s}%
\ddot{x}^{k}f^{l}\xi ^{m}\frac{\partial Q}{\partial \dot{x}^{i}}
\label{f5.4a}
\end{eqnarray}

Let us introduce designations 
\begin{eqnarray}
P_{i} &\equiv &P_{i}\left( \dot{x},\xi ,\dot{\xi},u\right) =-m\frac{\dot{x}%
_{i}}{\sqrt{\dot{x}^{s}\dot{x}_{s}}}-\hbar Q\varepsilon _{iklm}\ddot{x}%
^{k}u^{l}\xi ^{m}  \nonumber \\
&&-\frac{\hbar }{2}Q_{s}\ddot{x}^{s}\varepsilon _{iklm}\xi ^{m}\dot{x}%
^{k}u^{l}-\frac{\hbar }{2}Q\varepsilon _{iklm}\dot{\xi}^{m}\dot{x}^{k}u^{l}-%
\frac{\hbar }{2}Q_{i}\varepsilon _{sklm}\dot{x}^{s}\ddot{x}^{k}u^{l}\xi ^{m}
\label{f5.4}
\end{eqnarray}
where 
\begin{equation}
Q\equiv Q\left( \dot{x},u\right) =\frac{1}{\sqrt{\dot{x}^{s}\dot{x}_{s}}(%
\dot{x}^{l}u_{l}+\sqrt{\dot{x}^{l}\dot{x}_{l}})}  \label{f5.4b}
\end{equation}
\begin{equation}
Q_{i}\equiv Q_{i}\left( \dot{x},u\right) =\frac{\partial Q}{\partial \dot{x}%
^{i}}=-\frac{\dot{x}_{i}\left( \dot{x}^{l}u_{l}+2\sqrt{\dot{x}^{l}\dot{x}}%
_{l}\right) +\dot{x}_{l}\dot{x}^{l}u_{i}}{\left( \dot{x}^{s}\dot{x}%
_{s}\right) ^{3/2}(\dot{x}^{l}u_{l}+\sqrt{\dot{x}^{l}\dot{x}_{l}})^{2}}%
,\qquad i=0,1,2,3  \label{f5.5}
\end{equation}
Variation of the action (\ref{f5.1}) with respect to $x^{i}$ leads to the
dynamic equations 
\begin{equation}
\frac{d}{d\tau _{0}}P_{i}\left( \dot{x},\xi ,\dot{\xi},f\right) =0,\qquad
i=0,1,2,3  \label{f5.6}
\end{equation}
It means that the quantity $P_{i}$, defined by the equation (\ref{f5.4}) is
constant $P_{i}=$const., $i=0,1,2,3$.

Let us introduce new dynamic variables $u_{i}$ and $M$, defining them by the
relations 
\begin{equation}
Mu_{i}=-P_{i}\left( \dot{x},\xi ,\dot{\xi},u\right) ,\qquad u_{i}u^{i}-1=0
\label{f5.7}
\end{equation}
These constraints are added to the action (\ref{f5.1}) by means of the
Lagrangian multipliers $\lambda ^{i}$ and $\eta .$ Simultaneously we
identify the constant 4-vector $f^{k}$ with the 4-vector $u^{k}$, which is
dynamic variable and at the same time $u^{k}=$const in force of dynamic
equations (\ref{f5.6}) and designations (\ref{f5.7}). After identification $%
f^{k}=u^{k}$, the dynamic system ${\cal S}_{{\rm dc}}$ turns to a
relativistic dynamic system ${\cal S}_{{\rm dcr}}$, which is described by
the action 
\begin{eqnarray}
{\cal S}_{{\rm dcr}} &:&\;\;{\rm {\cal A}}_{{\rm dcr}}\left[ x,\xi
,u,M,\lambda ,\eta \right] =\int \left\{ -m\sqrt{\dot{x}^{i}\dot{x}_{i}}%
-\hbar \frac{\varepsilon _{iklm}\xi ^{i}\dot{\xi}^{k}u^{l}z^{m}}{2(1-\xi
^{s}z_{s})}\right.  \nonumber \\
&&-\left. \frac{\hbar }{2}Q\varepsilon _{iklm}\dot{x}^{i}\ddot{x}%
^{k}u^{l}z^{m}+\lambda ^{i}\left( Mu_{i}-P_{i}\right) +\eta \left(
u_{i}u^{i}-1\right) \right\} d\tau _{0}  \label{f5.8}
\end{eqnarray}
where $P_{i}$ and $Q$ are known functions (\ref{f5.4}), (\ref{f5.4b}) of
variables $\dot{x},u,\xi ,\dot{\xi}.$ Now the action (\ref{f5.8}) for the
dynamic system ${\cal S}_{{\rm dcr}}$ do not contain absolute objects, and
dynamic equations for ${\cal S}_{{\rm dcr}}$ appear to be compatible with
the relativity principles. Dynamic equations for ${\cal S}_{{\rm dcr}}$
coincide with dynamic equations ${\cal S}_{{\rm dc}}$ in the special case,
when the momentum $P_{i}$ of the dynamic system ${\cal S}_{{\rm dc}}$,
determined by the relations (\ref{f5.4a}), is chosen in such a way, that $%
P_{i}=-\sqrt{P_{s}P^{s}}f_{i}$. It means that the systems ${\cal S}_{{\rm dcr%
}}$ and ${\cal S}_{{\rm dc}}$ coincide in the nonrelativistic approximation,
when $f^{i}=\left\{ 1,0,0,0\right\} $. The procedure of elimination of the
absolute object $f^{i}$ and transformation of nonrelativistic dynamic system 
${\cal S}_{{\rm dc}}$ to the relativistic dynamic system ${\cal S}_{{\rm dcr}%
}$ will be referred to as a relativization of classical Dirac particle $%
{\cal S}_{{\rm dc}}$.

Note that the relativization procedure may be applied to the quantum Dirac
particle ${\cal S}_{{\rm D}}$ also. To make this, one should write the
action ${\cal A}_{{\rm Dqur}}$ for the statistical ensemble ${\cal E}_{{\rm %
Dqur}}$ of dynamic systems ${\cal S}_{{\rm dcr}}$

\begin{equation}
{\cal S}_{{\rm Dqur}}:\;\;{\cal A}_{{\rm Dqur}}\left[ x,\xi ,u,M,\lambda
,\eta \right] =\int L_{{\rm dcr}}d\tau _{0}d{\bf \tau }  \label{f5.9}
\end{equation}
where $L_{{\rm dcr}}$ is Lagrangian in the action (\ref{f5.8}), and
dependent dynamic variables $x,\xi ,u,M,\lambda ,\eta $ are considered to be
functions of independent variables $\tau =\left\{ \tau _{0},{\bf \tau }%
\right\} $. Thereafter the variables $x^{k}$ are considered to be
independent variables, and all other variables are considered to be
dependent ones. The action (\ref{f5.9}) transforms to the form 
\begin{equation}
{\cal S}_{{\rm Dqur}}:\;\;{\cal A}_{{\rm Dqur}}\left[ j,\xi ,u,M,\lambda
,\eta \right] =\int {\cal L}_{{\rm dcr}}d^{4}x  \label{f5.10}
\end{equation}
The action (\ref{f5.10}) differs from the action (\ref{f5.9}) as well as the
action (\ref{a5.13}) differs from (\ref{a5.18}). Thereafter one can apply
D-quantization to the action (\ref{f5.10}). It means that one should add
Lagrangian densities ${\cal L}_{{\rm q1}}$ and ${\cal L}_{{\rm q3}}$,
defined respectively by relations (\ref{c4.17}) and (\ref{a5.28}) (with $%
f^{k}$ substituted by $u^{k}$), to the Lagrangian density ${\cal L}_{{\rm dcr%
}}$ in (\ref{f5.10}). The Lagrangian densities ${\cal L}_{{\rm q1}}$ and $%
{\cal L}_{{\rm q3}}$ contain transversal derivatives and describe quantum
effects. One may return to description in terms of wave functions, but one
cannot be sure that dynamic equations appear to be linear in terms of wave
functions $\psi $.

\section{Solution of dynamic equations for ${\cal S}_{{\rm dcr}}$}

Having solved dynamic equations for ${\cal S}_{{\rm dcr}}$ at the coordinate
system, where 
\begin{equation}
u^{k}=\left\{ 1,0,0,0\right\}  \label{f6.1}
\end{equation}
one can obtain solutions for other values of $u^{k}=$const by means of
proper Lorentz transformations. At the values (\ref{f6.1}) dynamic equations
for ${\cal S}_{{\rm dcr}}$ coincide with dynamic equations for ${\cal S}_{%
{\rm dc}}$, provided they are written for the case $P_{i}=\left\{
P_{0},0,0,0\right\} ,$ $f^{k}=\left\{ 1,0,0,0\right\} .$ Thus, we shall
solve dynamic equations, derived from the action (\ref{b3.9}), which is
written for the case $f^{k}=\left\{ 1,0,0,0\right\} $.

Variation of the action (\ref{b3.9}) with respect to ${\bf x}$ gives the
dynamic equation 
\begin{equation}
\frac{d}{d\tau _{0}}\left( -m\frac{{\bf \dot{x}}}{\sqrt{\dot{x}^{s}\dot{x}%
_{s}}}+\frac{\hbar Q}{2}({\bf \xi }\times \ddot{{\bf x}})-\frac{\hbar }{2}%
\frac{\partial Q}{\partial {\bf \dot{x}}}(\dot{{\bf x}}\times \ddot{{\bf x}})%
{\bf \xi }+\frac{\hbar }{2}\frac{d}{d\tau _{0}}\left( Q({\bf \xi }\times 
{\bf \dot{x}})\right) \right) =0  \label{b6.1}
\end{equation}
where 
\begin{equation}
Q=Q\left( \dot{x}\right) =\left( \sqrt{\dot{x}^{s}\dot{x}_{s}}(\sqrt{\dot{x}%
^{s}\dot{x}_{s}}+\dot{x}^{0})\right) ^{-1},\qquad \dot{x}^{s}\dot{x}_{s}=%
\dot{x}_{0}^{2}-{\bf \dot{x}}^{2}  \label{b6.2}
\end{equation}
Varying the action (\ref{b3.9}) with respect to $x^{0}$, one obtains 
\begin{equation}
\frac{d}{d\tau _{0}}\left( m\frac{\dot{x}^{0}}{\sqrt{\dot{x}^{s}\dot{x}_{s}}}%
-\frac{\hbar }{2}\frac{\partial Q}{\partial \dot{x}^{0}}(\dot{{\bf x}}\times 
\ddot{{\bf x}}){\bf \xi }\right) =0  \label{b6.1a}
\end{equation}

Varying the action (\ref{b3.9}) with respect to ${\bf \xi ,}$ one should
take into account the side constraint ${\bf \xi }^{2}=1$. Setting 
\begin{equation}
\xi ^{\alpha }=\frac{\zeta ^{\alpha }}{\sqrt{{\bf \zeta }^{2}}},\qquad
\alpha =1,2,3  \label{b6.3}
\end{equation}
where ${\bf \zeta }$ is an arbitrary 3-pseudovector, one obtains 
\begin{equation}
\frac{\delta {\cal A}_{{\rm dc}}}{\delta \zeta ^{\mu }}=\frac{\delta {\cal A}%
_{{\rm dc}}}{\delta \xi ^{\alpha }}\frac{\delta \xi ^{\alpha }}{\delta \zeta
^{\mu }}=\frac{\delta {\cal A}_{{\rm dc}}}{\delta \xi ^{\alpha }}\frac{%
\delta ^{\alpha \mu }-\xi ^{\alpha }\xi ^{\mu }}{\sqrt{{\bf \zeta }^{2}}}=0
\label{b6.4}
\end{equation}
It means that there are only two independent components of dynamic equation (%
\ref{b6.4}). These components are orthogonal to 3-pseudovector ${\bf \xi }$
and can be obtained from equation $\delta {\cal A}_{{\rm dc}}/\delta \xi
^{\alpha }=0$ by means of vector product with ${\bf \xi }$. 
\begin{equation}
-\hbar {\frac{\left( \dot{{\bf \xi }}\times {\bf z}\right) \times {\bf \xi }%
}{2(1+{\bf z\xi })}+}\hbar \left( {-}\frac{d}{d\tau _{0}}{{\frac{({\bf \xi }%
\times {\bf z})}{2(1+{\bf z\xi })}}-\frac{(\dot{{\bf \xi }}\times {\bf \xi })%
{\bf z}}{2(1+{\bf z\xi })^{2}}{\bf z}}\right) \left( {\bf z}\times {\bf \xi }%
\right) +\hbar \frac{(\dot{{\bf x}}\times \ddot{{\bf x}})\times {\bf \xi }}{2%
}Q=0  \label{b6.5}
\end{equation}
After transformations this equation reduces to the equation (see Appendix\
C) 
\begin{equation}
{\bf \dot{\xi}}=-({\bf \dot{x}}\times {\bf \ddot{x}})\times {\bf \xi }Q,
\label{c7.1}
\end{equation}
which does not contain the vector ${\bf z}$. It means that ${\bf z}$
determines a fictitious direction in the space-time. Note that ${\bf z}$ in
the action (\ref{c4.15}) for the system ${\cal S}_{{\rm D}}$ is fictitious
also, because the term containing ${\bf z}$ is the same in both actions (\ref
{c4.15}) and (\ref{a5.13}) for ${\cal S}_{{\rm D}}$ and ${\cal S}_{{\rm Dqu}%
} $ respectively.

Let us choose the parameter $\tau _{0}$ in such a way, that 
\begin{equation}
\sqrt{\dot{x}^{s}\dot{x}_{s}}=\sqrt{\dot{x}_{0}^{2}-{\bf \dot{x}}^{2}}%
=1,\qquad \dot{x}_{0}=\sqrt{1+{\bf \dot{x}}^{2}}  \label{b6.10}
\end{equation}
Then, using the condition (\ref{b6.10}), one obtains 
\begin{equation}
Q=\frac{1}{1+\dot{x}_{0}},\qquad \frac{\partial Q}{\partial \dot{x}_{0}}%
=-1,\qquad \frac{\partial Q}{\partial {\bf \dot{x}}}=\frac{{\bf \dot{x}}%
\left( 2+\dot{x}_{0}\right) }{\left( 1+\dot{x}_{0}\right) ^{2}}
\label{b6.11}
\end{equation}

Integration of equation (\ref{b6.1a}) leads to

\begin{equation}
m\dot{x}_{0}+{\frac{\hbar }{2}}\left( {\bf \dot{x}}\times {\bf \ddot{x}}%
\right) {\bf \xi }=-p_{0}  \label{b6.7}
\end{equation}
where $p_{0}$ is an integration constant.

Eliminating $\left( {\bf \dot{x}}\times {\bf \ddot{x}}\right) {\bf \xi }$
from (\ref{b6.1}) by means of (\ref{b6.7}) and integrating it, one obtains
after simplification 
\begin{eqnarray}
&&\frac{d}{d\tau }\left( \frac{{\bf \dot{x}}}{\sqrt{1+\dot{x}_{0}}}\right)
\times {\bf \xi +}{\frac{1}{2}}\frac{{\bf \dot{x}}\times {\bf \dot{\xi}}}{%
\sqrt{1+\dot{x}_{0}}}  \nonumber \\
&=&\left( {\bf p}-m{\bf \dot{x}}\right) \frac{\sqrt{\left( 1+\dot{x}%
_{0}\right) }}{\hbar }+\frac{{\bf \dot{x}}\left( 2+\dot{x}_{0}\right) }{%
\hbar \left( \dot{x}_{0}+1\right) ^{3/2}}\left( p_{0}+m\dot{x}_{0}\right)
\label{b6.9}
\end{eqnarray}
where $p_{0},{\bf p}$ are integration constants, which are constant values
of 4-momentum $\left\{ P_{0},{\bf P}\right\} $ of the classical Dirac
particle defined by (\ref{f5.4a})

Let us set ${\bf p}=0$ and introduce new designations 
\begin{equation}
w_{0}=\frac{p_{0}}{m},\qquad \lambda =\frac{\hbar }{m}  \label{b6.12}
\end{equation}
\begin{equation}
{\bf y=}\frac{{\bf \dot{x}}}{\sqrt{1+\dot{x}_{0}}}{\bf =}\frac{{\bf \dot{x}}%
}{\sqrt{1+\sqrt{1+{\bf \dot{x}}^{2}}}},\;\;\;\;\;\;{\bf \dot{x}=y}\sqrt{%
\left( {\bf y}^{2}+2\right) }  \label{b6.13}
\end{equation}
\begin{equation}
\dot{x}_{0}=\sqrt{1+{\bf y}^{2}\left( {\bf y}^{2}+2\right) }={\bf y}^{2}+1
\label{b6.14}
\end{equation}

Dynamic equations (\ref{b6.9}), (\ref{b6.7}), (\ref{c7.1}) for dynamic
system ${\cal S}_{{\rm dc}}$ reduce to the form 
\begin{equation}
\lambda {\bf \dot{y}}\times \left( {\bf \xi +}{\frac{1}{2}}{\bf y}\left( 
{\bf y\xi }\right) \right) =-{\bf y}\left( \frac{1-w_{0}}{\left( {\bf y}%
^{2}+2\right) }-w_{0}\right)  \label{b7.1}
\end{equation}
\begin{equation}
\lambda {\bf \dot{y}}\left( {\bf y}\times {\bf \xi }\right) =2\left( 1-\frac{%
1-w_{0}}{\left( {\bf y}^{2}+2\right) }\right)  \label{b7.2}
\end{equation}
\begin{equation}
{\bf \dot{\xi}}=({\bf y\times \dot{y}})\times {\bf \xi }  \label{b7.3a}
\end{equation}
\begin{equation}
\dot{x}_{0}=\sqrt{1+{\bf y}^{2}\left( {\bf y}^{2}+2\right) }={\bf y}^{2}+1
\label{b7.3}
\end{equation}
It follows from (\ref{b7.1}) that ${\bf y}$ is orthogonal to ${\bf \dot{y}}$
and to ${\bf \xi +}{\frac{1}{2}}{\bf y}\left( {\bf y\xi }\right) $ 
\begin{equation}
\left( {\bf y\dot{y}}\right) =0,\qquad \left( {\bf y\xi }\right) \left( 1%
{\bf +}{\frac{1}{2}}{\bf y}^{2}\right) =0  \label{b7.4}
\end{equation}
Hence, 
\begin{equation}
{\bf y}^{2}=b=\text{const},\qquad \left( {\bf y\xi }\right) =0  \label{b7.5}
\end{equation}
and equation (\ref{b7.1}) has the form 
\begin{equation}
\lambda {\bf \dot{y}}\times {\bf \xi }=\left( -\frac{1-w_{0}}{b+2}%
+w_{0}\right) {\bf y}  \label{b7.6}
\end{equation}

Although it does not follow from (\ref{b7.1}) -- (\ref{b7.4}) directly that $%
\left( {\bf \dot{y}\xi }\right) =0$ and ${\bf \dot{\xi}}=0$, but special
investigation results, that ${\bf \xi }=$const, and one can find such a
coordinate system, where ${\bf \xi =}\left\{ 0,0,1\right\} $. In this
coordinate system 
\begin{equation}
\lambda \dot{y}_{2}=\left( -\frac{1-w_{0}}{b+2}+w_{0}\right) y_{1},\qquad
-\lambda \dot{y}_{1}=\left( -\frac{1-w_{0}}{b+2}+w_{0}\right) y_{2}
\label{b7.7}
\end{equation}
Solution of this system of equations has the form 
\begin{equation}
y_{1}=\sqrt{b}\cos \left( \omega \tau _{0}\right) ,\;\;\;y_{2}=\sqrt{b}\sin
\left( \omega \tau _{0}\right) ,  \label{b7.8}
\end{equation}
where 
\begin{equation}
\omega =\frac{1}{\lambda }\left( -\frac{1-w_{0}}{b+2}+w_{0}\right)
\label{b7.9}
\end{equation}

Substituting (\ref{b7.8}) into (\ref{b7.2}), one obtains 
\begin{equation}
-\left( -\frac{1-w_{0}}{b+2}+w_{0}\right) b=2\left( 1-\frac{1-w_{0}}{\left(
b+2\right) }\right)  \label{b7.10}
\end{equation}
It follows from (\ref{b7.10}) that integration constants $b$ and $w_{0}$ are
connected by the relation 
\begin{equation}
w_{0}=-\frac{1}{b+1}  \label{b7.11}
\end{equation}
As far as the vector ${\bf \xi }$ is orthogonal to both vectors ${\bf y}$
and ${\bf \dot{y}}$, the equation (\ref{b7.3a}) is satisfied by ${\bf \xi }=$%
const. Combining equations (\ref{b7.8}), (\ref{b7.3}) and (\ref{b6.13}), one
obtains 
\begin{equation}
\frac{d{\bf x}}{dt}={\bf y}\frac{\sqrt{b+2}}{b+1}=\left\{ \frac{\sqrt{%
b\left( b+2\right) }}{b+1}\cos \left( \Omega t\right) ,-\frac{\sqrt{b\left(
b+2\right) }}{b+1}\sin \left( \Omega t\right) ,0\right\}  \label{b7.12}
\end{equation}
where 
\begin{equation}
\Omega =\frac{\omega }{b+1}=\frac{2}{\lambda \left( b+1\right) ^{2}}
\label{b7.13}
\end{equation}
Integration of equation (\ref{b7.12}) results 
\begin{equation}
{\bf x}=\left\{ \lambda \frac{\left( b+1\right) \sqrt{b\left( b+2\right) }}{2%
}\sin \left( \Omega t\right) ,\lambda \frac{\left( b+1\right) \sqrt{b\left(
b+2\right) }}{2}\cos \left( \Omega t\right) ,0\right\}  \label{b7.14}
\end{equation}

In this special case the world line of the classical Dirac particle is a
helix. The total particle mass $m_{{\rm dcr}}=m|w_{0}|$, radius $a_{{\rm dcr}%
}$ of the helix and the particle velocity $v$ are determined by the
relations 
\begin{eqnarray}
m_{{\rm dcr}} &=&\frac{m}{b+1},\qquad a_{{\rm dcr}}=\hbar \frac{\left(
b+1\right) \sqrt{b\left( b+2\right) }}{2m}=\hbar \frac{\sqrt{b\left(
b+2\right) }}{2m_{{\rm dcr}}},  \label{b7.15} \\
v &=&\left| \frac{d{\bf x}}{dt}\right| =\frac{\sqrt{b\left( b+2\right) }}{%
\left( b+1\right) }  \label{b7.15.b}
\end{eqnarray}
Resolving the second equation (\ref{b7.15}) with respect to $b,$ one
expresses $b$ as a function of the radius $a_{{\rm dcr}}$%
\[
b=\frac{1}{\sqrt{2}}\sqrt{1+\sqrt{1+\zeta ^{2}}}-1,\qquad \zeta =4a_{{\rm dcr%
}}\frac{mc}{\hbar } 
\]
One can express the observable mass $m_{{\rm dcr}},$ velocity $v$ and the
angular velocity $\omega _{{\rm dcr}}$ as functions of the helix radius $a_{%
{\rm dcr}}.$ These relations have the following form 
\begin{equation}
m_{{\rm dcr}}=\frac{m\sqrt{2}}{\sqrt{\left( \sqrt{1+\zeta ^{2}}+1\right) }}=%
\frac{m}{\cosh \beta },\qquad \zeta =4a_{{\rm dcr}}\frac{mc}{\hbar }=\sinh
\left( 2\beta \right)  \label{b7.16}
\end{equation}
\begin{equation}
v=c\frac{\zeta }{\left( \sqrt{1+\zeta ^{2}}+1\right) }=c\tanh \beta ,\qquad
\omega _{{\rm dcr}}=\frac{4mc^{2}}{\hbar }\frac{1}{\left( \sqrt{1+\zeta ^{2}}%
+1\right) }=\frac{2mc^{2}}{\hbar \cosh ^{2}\beta }  \label{b7.17}
\end{equation}
Here $c$ is the speed of the light.

Note that $v$ is the Dirac particle velocity in the coordinate system, where
the particle momentum ${\bf p}=0$. It seems strange that the world line of a
free Dirac particle is a helix and the Dirac particle rotates. All this
looks as if there were two coupled particles rotating around their center of
mass, and we observe only one of them. Let us consider relativistic rotator
and test this hypothesis.

\section{ Relativistic rotator}

Rotator is a dynamic system ${\cal S}_{{\rm r}},$ consisting of two coupled
particles of mass $m_{0}$, which can rotate around their center of mass. The
distance between particles is to be constant, i.e. the particles are not to
vibrate. Rigid nonrelativistic rotator ${\cal S}_{{\rm nr}}$ is described by
the action 
\begin{equation}
{\cal S}_{{\rm nr}}:\qquad {\cal A}\left[ {\bf x}_{1},{\bf x}_{2},\mu \right]
=\int \left( \sum\limits_{k=1}^{k=2}\frac{m_{0}{\bf \dot{x}}_{k}^{2}}{2}+\mu
\left( \left( {\bf x}_{1}-{\bf x}_{2}\right) ^{2}-4a^{2}\right) \right) dt
\label{b8.46a}
\end{equation}
where $2a$ is the distance (length of string) between the particles. The
parameter $a$ is determined by the length of rigid coupling between two
particles. It does not depend on initial conditions.

If the coupling is elastic, the action for ${\cal S}_{{\rm nr}}$ should be
written in the form 
\begin{equation}
{\cal S}_{{\rm nr}}:\qquad {\cal A}\left[ {\bf x}_{1},{\bf x}_{2},\mu \right]
=\int \left( \sum\limits_{k=1}^{k=2}\frac{m_{0}{\bf \dot{x}}_{k}^{2}}{2}%
-U\left( \left( {\bf x}_{1}-{\bf x}_{2}\right) ^{2}\right) +\dot{\mu}\left( 
{\bf x}_{1}-{\bf x}_{2}\right) ^{2}\right) dt  \label{b8.46b}
\end{equation}
where $U$ is the potential energy, describing interaction energy between two
particles. This energy is constant for dynamic system (\ref{b8.46b}),
because of dynamic equation 
\begin{equation}
\frac{\delta {\cal A}}{\delta \mu }=-\frac{d}{dt}\left( {\bf x}_{1}-{\bf x}%
_{2}\right) ^{2}=0,\qquad \left( {\bf x}_{1}-{\bf x}_{2}\right) ^{2}=4a^{2}=%
\text{const}  \label{b8.46c}
\end{equation}
Evolution of variables ${\bf x}_{1}$, ${\bf x}_{2}$ does not depend on the
form of $U.$ Only $\mu $ depend on $U$. The last relation (\ref{b8.46c})
appears as integral of motion. Interesting only in evolution of ${\bf x}_{1} 
$, ${\bf x}_{2}$, one can omit the potential energy $U$ in the expression
for the action.

In the relativity theory a rigid coupling is impossible. Also there are no
reasons for introduction of a potential energy of interaction between the
particles, because in the relativity theory a long-range action is absent.
We are forced to choose another way.

Let us consider established relativistic motion of two particles of mass $%
m_{0}$, coupled between themselves by a massless elastic string. The
condition of established motion means that the particles move in such a way
that the length of the string does not change, and one may neglect degrees
of freedom, connected with the string. Mathematically it means, that there
is such a coordinate system $K$ (maybe, rotating), where particles are at
rest.

Let ${\cal L}_{1}$ and ${\cal L}_{2},$ be world lines of particles 
\begin{equation}
{\cal L}_{k}:\qquad x_{(k)}^{i}=x_{(k)}^{i}\left( \tau _{k}\right) ,\qquad
i=0,1,2,3;\;\;\;k=1,2  \label{b8.1}
\end{equation}
where $\tau _{k},\;\;k=1,2$ are parameters along these world lines. From
geometrical viewpoint the steady (established) motion of particles means
that any spacelike 3-plane ${\cal S}$, crossing ${\cal L}_{1}$ orthogonally,
crosses ${\cal L}_{2}$ also orthogonally. This circumstance permits one to
synchronize events on ${\cal L}_{1}$ and ${\cal L}_{2}$

Let parameters $\tau _{1}$ and $\tau _{2}$ are chosen in such a way that
they have the same value $\tau $ at points ${\cal L}_{1}\cap {\cal S}$ and $%
{\cal L}_{2}\cap {\cal S}$. The steady state conditions are written in the
form 
\begin{equation}
\frac{dx_{(1)}^{i}\left( \tau \right) }{d\tau }\left( x_{(1)i}\left( \tau
\right) -x_{(2)i}\left( \tau \right) \right) =\frac{dx_{(2)}^{i}\left( \tau
\right) }{d\tau }\left( x_{(1)i}\left( \tau \right) -x_{(2)i}\left( \tau
\right) \right) =0  \label{b8.2}
\end{equation}

Let us describe motion of the two particles by the action 
\begin{equation}
{\cal S}_{{\rm rr}}:\qquad {\cal A}_{{\rm rr}}\left[ x_{\left( 1\right)
},x_{\left( 2\right) }\right] =-\int \sum\limits_{k=1}^{k=2}m_{0}\sqrt{\dot{x%
}_{\left( k\right) }^{l}\dot{x}_{\left( k\right) l}}d\tau  \label{b8.3}
\end{equation}
where variables $x_{\left( k\right) }$ are considered to be functions of the
same parameter $\tau ,$ and the period denotes differentiation with respect
to $\tau $. World lines of particles are determined as extremals of the
functional (\ref{b8.3}) with side constraints (\ref{b8.2}).

It is convenient to introduce new variables\label{02} 
\begin{eqnarray}
X^{k} &=&\frac{1}{2}\left( x_{\left( 1\right) }^{k}+x_{\left( 2\right)
}^{k}\right) ,\;\;\;x^{k}=\frac{1}{2}\left( x_{\left( 1\right)
}^{k}-x_{\left( 2\right) }^{k}\right) ,  \label{b8.5} \\
x_{\left( 1\right) }^{k} &=&X^{k}+x^{k},\;\;\;\;\;x_{\left( 2\right)
}^{k}=X^{k}-x^{k},  \label{b8.6}
\end{eqnarray}
and rewrite the constraints (\ref{b8.2}) in the equivalent form 
\begin{equation}
\frac{d}{d\tau }\left( x^{k}x_{k}\right) =0,\qquad \dot{X}^{k}x_{k}=0
\label{b8.7}
\end{equation}
Here the first condition describes constancy of the string length. This
condition is used in the nonrelativistic case also. The second condition is
the synchronization condition, which is possible only for established motion
of two rotator particles. This condition is not used in the nonrelativistic
case in explicit form.

Let us add conditions (\ref{b8.7}) to the action by means of Lagrangian
multipliers $\mu $ and $\nu .$ Then one obtains the following expression for
the action 
\begin{equation}
{\cal S}_{{\rm rr}}:\qquad {\cal A}_{{\rm rr}}\left[ X,x,\mu ,\nu \right]
=\int \left\{ -m_{0}R-m_{0}r+\dot{\mu}x^{k}x_{k}+\nu \dot{X}%
^{k}x_{k}\right\} d\tau  \label{b8.8}
\end{equation}
where 
\begin{equation}
R=\sqrt{\left( \dot{X}^{k}+\dot{x}^{k}\right) \left( \dot{X}_{k}+\dot{x}%
_{k}\right) },\qquad r=\sqrt{\left( \dot{X}^{k}-\dot{x}^{k}\right) \left( 
\dot{X}_{k}-\dot{x}_{k}\right) }  \label{b8.9}
\end{equation}

Momenta are expressed as follows 
\begin{eqnarray}
P_{k} &=&\frac{\partial L}{\partial \dot{X}^{k}}=-\frac{m_{0}}{R}\left( \dot{%
X}_{k}+\dot{x}_{k}\right) -\frac{m_{0}}{r}\left( \dot{X}_{k}-\dot{x}%
_{k}\right) +\nu x_{k},  \label{b8.10} \\
p_{k} &=&\frac{\partial L}{\partial x^{k}}=-\frac{m_{0}}{R}\left( \dot{X}%
_{k}+\dot{x}_{k}\right) +\frac{m_{0}}{r}\left( \dot{X}_{k}-\dot{x}%
_{k}\right) ,  \label{b8.11}
\end{eqnarray}

The action (\ref{b8.8}) is invariant with respect to transformation $\tau
\rightarrow f\left( \tau \right) .$ Using this, one chooses the parameter $%
\tau $ in such a way that $R+r=1.$ Introducing designation 
\begin{equation}
\beta =R-r,\qquad R+r=1,\qquad R=\frac{1+\beta }{2},\qquad r=\frac{1-\beta }{%
2},  \label{b8.11a}
\end{equation}
and resolving relations (\ref{b8.10}), (\ref{b8.11}) with respect to $\dot{X}
$ and $\dot{x}$, one obtains dynamic equations 
\begin{eqnarray}
\dot{X}_{k} &=&-\frac{1}{4m_{0}}\left( P_{k}+\beta p_{k}-\nu x_{k}\right)
\label{b8.12} \\
\dot{x}_{k} &=&-\frac{1}{4m_{0}}\left( \beta P_{k}+p_{k}-\beta \nu
x_{k}\right)  \label{b8.13}
\end{eqnarray}
Variation with respect to $X^{k},x^{k},\mu ,\nu \;$gives respectively 
\begin{eqnarray}
-\dot{P}_{k} &=&0,\qquad \dot{p}_{k}=2\dot{\mu}x_{k}+\nu \dot{X}_{k}
\label{b8.14} \\
2\left( \dot{x}^{k}x_{k}\right) &=&0,\qquad \dot{X}^{k}x_{k}=0,
\label{b8.15}
\end{eqnarray}
Equations (\ref{b8.12}) -- (\ref{b8.15}) form the complete system of dynamic
equations for determination of variables $X^{k},P_{k},x^{k},p_{k},\mu ,\nu $
as functions of parameter $\tau $.

It follows from the first equations (\ref{b8.14}) and (\ref{b8.15}) 
\begin{equation}
P_{k}=\text{const, \qquad }k=0,1,2,3,\qquad x^{k}x_{k}=-a^{2}=\text{const}
\label{b8.16}
\end{equation}
Convoluting equations (\ref{b8.12}), (\ref{b8.13}) and the second equation (%
\ref{b8.14}) with $x^{k}$ and using equations (\ref{b8.15}), one obtains the
relations 
\begin{equation}
\nu =-\frac{P_{k}x^{k}}{a^{2}},\qquad p_{k}x^{k}=0,\;\;\;\;\dot{\mu}=-\frac{%
x^{k}\dot{p}_{k}}{2a^{2}}  \label{b8.17}
\end{equation}
\begin{equation}
\dot{p}_{s}x^{s}=-p_{s}\dot{x}^{s}=\frac{p^{k}}{4m_{0}}\left( \beta
P_{k}+p_{k}\right)  \label{b8.17a}
\end{equation}
$\dot{\mu}$ is expressed as follows\ 
\begin{equation}
\dot{\mu}=-\frac{\dot{p}_{s}x^{s}}{2a^{2}}=-\frac{p^{k}}{8m_{0}a^{2}}\left(
\beta P_{k}+p_{k}\right)  \label{b8.18b}
\end{equation}

Let us now substitute equations (\ref{b8.12}), (\ref{b8.13}) in relations (%
\ref{b8.9}). Using relations (\ref{b8.17}), one obtains the conditions 
\begin{eqnarray*}
\left( 2m_{0}\right) ^{2} &=&\left( P_{k}+p_{k}\right) \left(
P^{k}+p^{k}\right) +\nu ^{2}a^{2} \\
\left( 2m_{0}\right) ^{2} &=&\left( P_{k}-p_{k}\right) \left(
P^{k}-p^{k}\right) +\nu ^{2}a^{2}
\end{eqnarray*}
Combining them, one obtains 
\begin{equation}
p^{k}p_{k}=-\left( P^{k}P_{k}-4m_{0}^{2}\right) -a^{2}\nu ^{2},\qquad
P_{k}p^{k}=0  \label{b8.18}
\end{equation}
Let us substitute (\ref{b8.18b}) into the second equation (\ref{b8.14}) and
take into account constraint (\ref{b8.18}) and dynamic equation (\ref{b8.12}%
). The second equation (\ref{b8.14}) reduces to 
\begin{equation}
\dot{p}_{k}=-x_{k}\frac{1}{4m_{0}a^{2}}\left( 4m_{0}^{2}-P^{s}P_{s}\right) -%
\frac{\nu }{4m_{0}}\left( P_{k}+\beta p_{k}\right)  \label{b8.21}
\end{equation}

Convoluting (\ref{b8.21}) with $p^{k}$ and taking into account (\ref{b8.17}%
), one obtains 
\begin{equation}
p^{k}\dot{p}_{k}=-\beta \nu \frac{p^{k}p_{k}}{4m_{0}}  \label{b8.22}
\end{equation}

Differentiating the first equation (\ref{b8.18}) and taking into account (%
\ref{b8.17}), (\ref{b8.13}), one obtains\ 
\begin{equation}
p^{k}\dot{p}_{k}=-a^{2}\nu \dot{\nu}=-\frac{\beta \nu }{4m_{0}}\left(
P_{k}P^{k}+\nu ^{2}a^{2}\right)  \label{b8.22a}
\end{equation}
Comparing relations (\ref{b8.22}), (\ref{b8.22a}) and taking into account (%
\ref{b8.18}), one concludes that 
\begin{equation}
\beta \nu \left( p_{k}p^{k}-2m_{0}^{2}\right) =0  \label{b8.25}
\end{equation}
If $p^{k}p_{k}=2m_{0}^{2}=$const, then it follows from (\ref{b8.22}) that 
\begin{equation}
\nu \beta =0  \label{b8.50}
\end{equation}
and constraint (\ref{b8.25}) reduces to (\ref{b8.50})

It follows from (\ref{b8.50}), that 
\begin{equation}
\beta =0\wedge \nu =0  \label{b8.25a}
\end{equation}
To show this, let us introduce spacelike vector 
\begin{equation}
\zeta _{i}=\varepsilon _{iklm}x^{k}p^{l}P^{m}.  \label{b8.33}
\end{equation}
It follows from dynamic equations (\ref{b8.13}), (\ref{b8.21}) that 
\begin{equation}
\dot{\zeta}_{i}=\varepsilon _{iklm}\dot{x}^{k}p^{l}P^{m}+\varepsilon
_{iklm}x^{k}\dot{p}^{l}P^{m}=0  \label{b8.34}
\end{equation}
Let 
\begin{equation}
\xi _{i}=\frac{\zeta _{i}}{\sqrt{-\zeta _{k}\zeta ^{k}}},\qquad \xi _{i}\xi
^{i}=-1  \label{b8.35}
\end{equation}
Thus, $\xi _{i}$ is a unit constant spacelike vector, which is orthogonal to 
$P_{k},p_{k},x_{k}$. 4-vector $p_{k}$ is orthogonal to vectors $P_{i}$ and $%
\xi _{i}$.

Let us choose such a coordinate system, where 
\begin{equation}
P_{i}=\left\{ P_{0},0,0,0\right\} ,\;\;\;\;\xi _{i}=\left\{ 0,0,0,1\right\} ,
\label{b8.36}
\end{equation}
Then 
\begin{equation}
x_{i}=\left\{ x_{0},x_{1},x_{2},0\right\} ,\qquad p_{i}=\left\{
0,p_{1},p_{2},0\right\} ,  \label{b8.37}
\end{equation}
Let $\nu =0$ be a solution of (\ref{b8.50}). For $k=0$ the equations (\ref
{b8.13}), (\ref{b8.21}) have respectively the form

\begin{equation}
\dot{x}_{0}=-\frac{1}{4m_{0}}\beta P_{0},\qquad \dot{p}_{0}=0=\frac{%
P_{0}^{2}-4m_{0}^{2}}{4m_{0}a^{2}}x_{0}  \label{b8.38a}
\end{equation}
It follows from (\ref{b8.38a}) that $x_{0}=0$, and $\beta =0$.

Let now $\beta =0.$ Then for $k=0$ the equations (\ref{b8.13}), (\ref{b8.21}%
) have respectively the form 
\begin{equation}
\dot{x}_{0}=0,\qquad 0=\frac{P_{0}^{2}-4m_{0}^{2}}{4m_{0}a^{2}}x_{0}-\frac{1%
}{4m_{0}}\nu P_{0}  \label{b8.38b}
\end{equation}

It follows from (\ref{b8.38b}) that $x_{0}=$const, 
\begin{equation}
\nu =\frac{P_{0}^{2}-4m_{0}^{2}}{P_{0}a^{2}}x_{0}=\text{const}  \label{b8.51}
\end{equation}
The condition (\ref{b8.51}) appears to be compatible with (\ref{b8.18}) and
the second constraint (\ref{b8.16}), provided $\nu =0$ or $m_{0}=0.$ As far
as $m_{0}\neq 0$, the relation (\ref{b8.25a}) takes place and $x_{0}=0$. The
dynamic equations (\ref{b8.13}), (\ref{b8.21}) take the form

\begin{equation}
\dot{x}_{\alpha }=-\frac{p_{\alpha }}{4m_{0}},\qquad \dot{p}_{\alpha }=\frac{%
P_{0}^{2}-4m_{0}^{2}}{4m_{0}a^{2}}x_{\alpha },\qquad \alpha =1,2
\label{b8.55}
\end{equation}

Solution of dynamic equations (\ref{b8.55}), (\ref{b8.12}) is written in the
form 
\begin{eqnarray}
x^{i} &=&\left\{ 0,\;a\cos \left( \omega \tau +\phi \right) ,\;a\sin \left(
\omega \tau +\phi \right) ,\;0\right\}  \label{b8.39} \\
p_{i} &=&\left\{ 0,\;-4am_{0}\omega \sin \left( \omega \tau +\phi \right)
,\;4am_{0}\omega \cos \left( \omega \tau +\phi \right) ,\;0\right\}
\label{b8.40} \\
X^{k} &=&\left\{ -\frac{P_{0}}{4m_{0}}\tau ,0,0,0\right\} ,\;\;\;X_{0}=t=-%
\frac{P_{0}}{4m_{0}}\tau  \label{b8.41}
\end{eqnarray}
where 
\begin{equation}
\omega =\frac{\sqrt{P_{0}^{2}-4m_{0}^{2}}}{4m_{0}a},  \label{b8.42}
\end{equation}
and $\phi $ is an arbitrary constant.

Let us substitute the independent variable $\tau $ by $t=-\frac{P_{0}}{4m_{0}%
}\tau $. Then one obtains 
\begin{equation}
\omega \tau =\omega _{0}t,\qquad \omega _{0}=-\frac{\sqrt{%
P_{0}^{2}-4m_{0}^{2}}}{aP_{0}}  \label{b8.43}
\end{equation}
\begin{eqnarray}
x_{(1)}^{i} &=&\left\{ t,\;a\cos \left( \omega _{0}t+\phi \right) ,\;a\sin
\left( \omega _{0}t+\phi \right) ,\;0\right\}  \label{b8.44} \\
x_{(2)}^{i} &=&\left\{ t,\;-a\cos \left( \omega _{0}t+\phi \right) ,\;-a\sin
\left( \omega _{0}t+\phi \right) ,\;0\right\}  \label{b8.45}
\end{eqnarray}

One can see that both world lines ${\cal L}_{1}$ and ${\cal L}_{2}$ are
helixes. They describe rotation of two particles around their common center
of mass along a circle of radius $a$. Such a dynamic system ${\cal S}_{{\rm %
rr}}$ may be qualified as a relativistic rotator. This rotator may be
described by the relative mass increase $\gamma =\left( M-2m_{0}\right)
/2m_{0}$, where 
\begin{equation}
\gamma =\frac{\left( M-2m_{0}\right) }{2m_{0}}=\frac{1}{\sqrt{1-v^{2}}}-1=%
\frac{v^{2}}{\sqrt{1-v^{2}}\left( \sqrt{1-v^{2}}+1\right) }  \label{b8.76}
\end{equation}
Here $M=\sqrt{P_{i}P^{i}}$ is the total mass of the rotator, and $v=a\omega
_{0}$ is the velocity of a particle in the coordinate system, where center
of mass is at rest. $\gamma $ is a part of total mass, conditioned by
rotation.

The distance $2a$ between particles appears as an integration constant. In
the solution (\ref{b8.44}) the radius $a$ and angular frequency $\omega _{0}$
of rotation are independent integration constants.

For any real rotator these quantities cannot be independent. Elastic
properties of the coupling between two particles determine the relative mass
increase $\gamma $. The elastic properties can be described by the rigidity
function $\gamma =$ $f_{{\rm r}}\left( a\right) $, which determines relation
between quantities $a$ and $\omega _{0}$, or between $a$ and $\gamma $ for
any real rotator. For nonrelativistic rotator the rigidity function $f_{{\rm %
r}}\left( a\right) $ is connected with the potential energy $U\left(
a\right) $ of elastic coupling by means of relation 
\[
f_{{\rm r}}\left( a\right) =\frac{a}{mc^{2}}\frac{\partial U\left( a\right) 
}{\partial a} 
\]
In the relativistic case one cannot introduce potential energy of elastic
coupling. It is replaced by the rigidity function $f_{{\rm r}}\left(
a\right) $.

The dynamic system ${\cal S}_{{\rm dcr}}$, where angular velocity is coupled
with the radius of helix, is a special case of relativistic rotator ${\cal S}%
_{{\rm rr}}$. To show this, let us compare relations (\ref{b8.43}), (\ref
{b8.44}) with relations (\ref{b7.13}), (\ref{b7.14}) and identify the
quantities $\omega _{0}$, $M$, $a$ of dynamic system ${\cal S}_{{\rm rr}}$
respectively with $\omega _{{\rm dcr}}$, $m_{{\rm dcr}}$, $a_{{\rm dcr}}$ of
dynamic system ${\cal S}_{{\rm dcr}}$. One derives 
\begin{eqnarray}
\frac{\sqrt{M^{2}-4m_{0}^{2}}}{aM} &=&\frac{4m}{\hbar }\frac{1}{\left( \sqrt{%
1+\zeta ^{2}}+1\right) },\qquad M=\frac{m\sqrt{2}}{\sqrt{\left( \sqrt{%
1+\zeta ^{2}}+1\right) }}  \label{b8.71} \\
a &=&a_{{\rm dcr}},\qquad \zeta =4a_{{\rm dcr}}\frac{m}{\hbar },\qquad c=1
\label{b8.72}
\end{eqnarray}

Resolving relations (\ref{b8.71})\label{00} with respect to variables $M$
and $m_{0}$, one obtains all quantities of the relativistic rotator ${\cal S}%
_{{\rm rr}}$ in terms of parameters $\zeta =4\hbar a_{{\rm dcr}}/m$ and $m$
of the Dirac particle ${\cal S}_{{\rm dcr}}$%
\begin{equation}
M=m_{{\rm dcr}}=\frac{m\sqrt{2}}{\sqrt{\left( \sqrt{1+\zeta ^{2}}+1\right) }}%
,\qquad m_{0}=\frac{m}{\left( \sqrt{1+\zeta ^{2}}+1\right) }  \label{b8.73}
\end{equation}
One can express parameters $m,$ $m_{{\rm dcr}},$ $\omega _{{\rm dcr}},$ $v_{%
{\rm dcr}}$ of ${\cal S}_{{\rm dcr}}$ in terms of parameters $v=\zeta
_{0}=4am_{0}/\hbar $, $m_{0}$ of the relativistic rotator ${\cal S}_{{\rm rr}%
}$. After some calculations one obtains 
\begin{equation}
m=2m_{0}\frac{1}{1-v^{2}},\qquad m_{{\rm dcr}}=\frac{2m_{0}}{\sqrt{1-v^{2}}}%
,\qquad \omega _{{\rm dcr}}=\frac{4m_{0}c^{2}}{\hbar },  \label{b8.74}
\end{equation}
where 
\begin{equation}
v=\zeta _{0}=4a\frac{m_{0}c^2}{\hbar }  \label{b8.75}
\end{equation}
The relativistic angular momentum $A$ and magnetic moment $\mu _{0}$ have
the form

\[
A=2m_{0}\frac{av}{\sqrt{1-v^{2}}},\qquad \mu _{0}=\frac{eav}{2\sqrt{1-v^{2}}}
\]
\[
\frac{\mu _{0}}{A}=\frac{e}{4m_{0}}=\frac{e}{2m_{{\rm dcr}}}\sqrt{1-v^{2}} 
\]
The rigidity function $\gamma =$ $f_{{\rm r}}\left( a_{{\rm dcr}}\right) $
is described by relations (\ref{b8.76}). As it follows from relations (\ref
{b8.74}), (\ref{b8.75}) the rigidity function has the form 
\begin{equation}
\gamma =f_{{\rm r}}\left( a\right) =\frac{\hbar }{\sqrt{\hbar ^{2}-\left(
4am_{0}c\right) ^{2}}}-1  \label{b8.80a}
\end{equation}
It follows from (\ref{b8.80a}) that $a\leq \frac{\hbar }{4m_{0}c}$.
Apparently, this points to quantum origin of interaction between the rotator
particles.

\section{Discussion}

It is a common practice to consider the Dirac particle to be a simple
pointlike construction, which has its proper characteristics such as mass,
charge, spin and magnetic moment. Spin and magnetic moment seem to be
extraneous for pointlike construction. Directions of momentum and velocity
do not coincide for Dirac particle. It also seems rather strange for
pointlike structure. The classical Dirac particle has ten degrees of freedom
in the sense, that integration of dynamic equations leads to appearance of
ten integration constants. It is too much for pointlike structure.

Dynamic analysis of dynamic system ${\cal S}_{{\rm D}}$ shows that the Dirac
particle is not a pointlike structure. It is a more complicated structure,
consisting of two pointlike coupled particles. Such properties as spin and
magnetic moment are quite reasonable for such a dynamic system. Discrepancy
between the direction of velocity and that of momentum is explained also, if
one compares alternating velocity of a particle with the constant momentum
of the whole rotator. The number of degrees of freedom is explained freely
also by existence of two coupled particles. Six external degrees of freedom,
describing motion of the particle as a whole, are relativistic in the sense
that the global velocity is less, than the speed of light. If one neglects
the internal degrees of freedom, described by two last terms in the action (%
\ref{a5.18}), the dynamic system ${\cal S}_{{\rm dc}}$ becomes relativistic.
The nonrelativistic character of the dynamic system ${\cal S}_{{\rm dc}}$ is
connected with incorrect description of internal degrees of freedom. The
dynamic system ${\cal S}_{{\rm dc}}$ can be made relativistic, if one
supposes that ${\cal S}_{{\rm dc}}$ is a special case of ${\cal S}_{{\rm rr}%
} $ with the rigidity function $\gamma =$ $f_{{\rm r}}\left( a\right) $,
described by the relation (\ref{b8.80a}). The rigidity function $\gamma =$ $%
f_{{\rm r}}\left( a\right) $, is a relativistically invariant characteristic
of the relativistic rotator ${\cal S}_{{\rm rr}}$.

Explaining freely all strange properties of the Dirac particle, the
two-particle model poses a very important question. What is an origin of the
coupling between two particles? Appearance of quantum constant $\hbar $ in
the relation (\ref{b8.80a}), describing character of coupling, shows, that
the coupling origin is, apparently, quantum.

Conventional quantum mechanics failed to discover internal structure of
Dirac particle. Methods of QM are too rough. Internal structure of Dirac
particle is discovered by dynamic methods of MCQP which appear to be more
subtle. Internal structure of the Dirac particle is of no importance for
calculations of stationary atom states, because characteristic energies are
too small, and internal degrees of freedom of the Dirac particle are not
excited. But the Dirac particle structure may appear to be important at
investigation of elementary particles structure, because characteristic
energies are large enough, and internal degrees of freedom can be excited.

\newpage \renewcommand{\theequation}{\Alph{section}.\arabic{equation}} %
\renewcommand{\thesection}{\Alph{section}} \setcounter{section}{0} %
\centerline{\Large \bf Mathematical Appendices}

\section{Calculation of Lagrangian}

Let us calculate the expression 
\begin{equation}
{\frac{i}{2}}\hbar \bar{\psi}\gamma ^{l}\partial _{l}\psi +\text{h.c}%
=F_{1}+F_{2}+F_{3}+F_{4}  \label{c8.1}
\end{equation}
where the following designations are used 
\begin{equation}
F_{1}={\frac{i}{2}}\hbar \psi ^{\ast }\left( \left( \partial _{0}-i\gamma
_{5}{\bf \sigma \nabla }\right) i\varphi \right) \psi +\text{h.c.}
\label{c8.2}
\end{equation}
\begin{equation}
F_{2}={\frac{i}{2}}\hbar \psi ^{\ast }\left( \left( \partial _{0}-i\gamma
_{5}{\bf \sigma \nabla }\right) \left( \frac{1}{2}\gamma _{5}\kappa \right)
\right) \psi +\text{h.c.}  \label{c8.3}
\end{equation}
\begin{equation}
F_{3}=+\frac{i}{2}\hbar A^{2}\Pi \left( \left( {\bf \sigma n}\right)
e^{-i\gamma _{5}\bsigma\bmeta}\left( {\bf \sigma n}\right) \right) e^{-\frac{%
i\pi }{2}\bsigma {\bf n}}(\partial _{0}-i\gamma _{5}{\bf \sigma \nabla })(e^{%
\frac{i\pi }{2}\bsigma {\bf n}})\Pi +\hbox{h.c.}  \label{c8.4}
\end{equation}
\begin{equation}
F_{4}={\frac{i}{2}}\hbar A^{2}\Pi e^{-{\frac{i}{2}}\gamma _{5}{\bf \Sigma }%
\bmeta}(\partial _{0}-i\gamma _{5}{\bf \Sigma \nabla })e^{-{\frac{i}{2}}%
\gamma _{5}{\bf \Sigma }\bmeta}\Pi +\hbox{h.c.}  \label{c8.5}
\end{equation}
In the last relation the matrix ${\bf \Sigma }$ is not differentiated.

Using definitions of $j^{l}$ and $S^{l}$, the expression $F_{1}$ and $F_{2}$
reduce to the form 
\begin{equation}
F_{1}={\frac{i}{2}}\hbar \psi ^{\ast }\left( \left( \partial _{0}-i\gamma
_{5}{\bf \sigma \nabla }\right) i\varphi \right) \psi +\text{h.c.}%
=-j^{l}\partial _{l}\varphi \Pi  \label{b5.5}
\end{equation}
\begin{eqnarray*}
F_{2} &=&{\frac{i}{2}}\hbar \psi ^{\ast }\left( \left( \partial _{0}-i\gamma
_{5}{\bf \sigma \nabla }\right) \left( \frac{1}{2}\gamma _{5}\kappa \right)
\right) \psi +\text{h.c} \\
&=&{\frac{i}{2}}\hbar \psi ^{\ast }\gamma _{5}\gamma ^{l}\partial _{l}\left( 
\frac{1}{2}\gamma _{5}\kappa \right) \psi +\text{h.c}
\end{eqnarray*}
\begin{equation}
F_{2}=-{\frac{1}{2}}\hbar S^{l}\partial _{l}\kappa \Pi  \label{b5.6}
\end{equation}

\begin{eqnarray*}
F_{3} &=&\frac{i}{2}\hbar A^{2}\Pi \left( e^{-\frac{i\pi }{2}\bsigma {\bf n}%
}e^{-i\gamma _{5}\bsigma\bmeta}e^{\frac{i\pi }{2}\bsigma {\bf n}}\right) e^{-%
\frac{i\pi }{2}\bsigma {\bf n}}(\partial _{0}-i\gamma _{5}{\bf \sigma }%
\nabla )(e^{\frac{i\pi }{2}\bsigma {\bf n}})\Pi +\hbox{h.c.} \\
&=&\frac{i}{2}\hbar j^{l}\Pi \sigma _{\alpha }\sigma _{\beta }n^{\alpha
}\partial _{l}n^{\beta }\Pi +\hbox{h.c.}=\frac{i}{2}\hbar j^{l}n^{\alpha
}\partial _{l}n^{\beta }\Pi \left( \delta _{\alpha \beta }+i\varepsilon
_{\alpha \beta \gamma }\sigma _{\gamma }\right) \Pi +\hbox{h.c.} \\
&=&\frac{i}{2}\hbar j^{l}\left( n^{\alpha }\partial _{l}n^{\alpha
}+i\varepsilon _{\alpha \beta \gamma }n^{\alpha }\partial _{l}n^{\beta
}z^{\gamma }\right) \Pi +\hbox{h.c.}
\end{eqnarray*}
As far as ${\bf n}^{2}=1$, one obtains 
\begin{equation}
n^{\alpha }\partial _{l}n^{\alpha }=0  \label{b5.8}
\end{equation}
Besides it follows from (\ref{a3.21}) that 
\begin{equation}
{\bf n}=\frac{{\bf \sigma }+{\bf z}}{\sqrt{2\left( 1+{\bf \sigma z}\right) }}
\label{b5.9}
\end{equation}

Then 
\begin{equation}
F_{3}=-\hbar j^{l}\left( \varepsilon _{\alpha \beta \gamma }n^{\alpha
}\partial _{l}n^{\beta }z^{\gamma }\right) \Pi =-\frac{\hbar j^{l}}{2\left(
1+{\bf \xi z}\right) }\varepsilon _{\alpha \beta \gamma }\xi ^{\alpha
}\partial _{l}\xi ^{\beta }z^{\gamma }\Pi  \label{c8.9}
\end{equation}
Calculation of $F_{4}$ leads to the following result 
\begin{eqnarray*}
F_{4} &=&\frac{i}{2}\hbar A^{2}\Pi e^{-\frac{i}{2}\gamma _{5}{\bf \Sigma }%
\bmeta}(\partial _{0}-i\gamma _{5}{\bf \Sigma \nabla })e^{-\frac{i}{2}\gamma
_{5}{\bf \Sigma }\bmeta}\Pi +\hbox{h.c. } \\
&=&\frac{i}{2}\hbar A^{2}\Pi \left( \cosh \frac{\eta }{2}-i\gamma
_{5}v^{\alpha }\Sigma _{\alpha }\sinh \frac{\eta }{2}\right) \\
&&\times (\partial _{0}-i\gamma _{5}{\bf \Sigma \nabla })\left( \cosh \frac{%
\eta }{2}-i\gamma _{5}v^{\alpha }\Sigma _{\alpha }\sinh \frac{\eta }{2}%
\right) \Pi +\hbox{h.c.
} \\
&=&\frac{i}{2}\hbar A^{2}\Pi \left( \cosh \frac{\eta }{2}\sinh \frac{\eta }{2%
}\partial _{0}\eta +\sinh ^{2}\frac{\eta }{2}v^{\alpha }\partial
_{0}v^{\beta }\Sigma _{\alpha }\Sigma _{\beta }\right) \Pi +\hbox{h.c. } \\
&&+\frac{i}{2}\hbar A^{2}\Pi \left( \cosh ^{2}\frac{\eta }{2}\Sigma _{\alpha
}\Sigma _{\beta }v^{\beta }+\sinh ^{2}\frac{\eta }{2}\Sigma _{\beta }\Sigma
_{\alpha }v^{\beta }\right) \partial _{\alpha }\frac{\eta }{2}\Pi +%
\hbox{h.c. } \\
&&+\frac{i}{2}\hbar A^{2}\Pi \cosh \frac{\eta }{2}\sinh \frac{\eta }{2}%
\Sigma _{\alpha }\Sigma _{\beta }\partial _{\alpha }v^{\beta }\Pi +%
\hbox{h.c. }
\end{eqnarray*}
\begin{eqnarray*}
F_{4} &=&\frac{i}{2}\hbar A^{2}\Pi \left( \frac{1}{2}\sinh \eta \partial
_{0}\eta +\sinh ^{2}\frac{\eta }{2}v^{\alpha }\partial _{0}v^{\beta
}i\varepsilon _{\alpha \beta \gamma }\Sigma _{\gamma }\right) \Pi +%
\hbox{h.c. }+ \\
&&+\frac{i}{2}\hbar A^{2}\Pi \left( \cosh \eta v^{\alpha }+i\varepsilon
_{\beta \alpha \gamma }v^{\beta }\Sigma _{\gamma }\right) \partial _{\alpha }%
\frac{\eta }{2}\Pi +\hbox{h.c. } \\
&&+\frac{i}{4}\hbar A^{2}\Pi \sinh \eta \left( \partial _{\alpha }v^{\alpha
}+i\varepsilon _{\alpha \beta \gamma }\partial _{\alpha }v^{\beta }\Sigma
_{\gamma }\right) \Pi +\hbox{h.c. }
\end{eqnarray*}
\begin{eqnarray*}
F_{4} &=&-\hbar A^{2}\left( \sinh ^{2}\frac{\eta }{2}v^{\alpha }\partial
_{0}v^{\beta }\varepsilon _{\alpha \beta \gamma }\xi ^{\gamma }\right) \Pi \\
&&-\frac{1}{2}\hbar A^{2}\Pi \left( \varepsilon _{\beta \alpha \gamma
}v^{\beta }\xi ^{\gamma }\partial _{\alpha }\eta +\sinh \eta \varepsilon
_{\alpha \beta \gamma }\partial _{\alpha }v^{\beta }\xi ^{\gamma }\right) \Pi
\end{eqnarray*}
\begin{equation}
F_{4}=-{\frac{1}{2}}\hbar A^{2}\varepsilon _{\alpha \beta \gamma }\left(
\partial _{\alpha }\eta v^{\beta }+\sinh \eta \partial _{\alpha }v^{\beta
}+2\sinh ^{2}({\frac{\eta }{2}})v^{\alpha }\partial _{0}v^{\beta }\right)
\xi ^{\gamma }\Pi  \label{c8.10}
\end{equation}

\section{Transformation of Lagrangian to covariant form}

Let us show that the expression (\ref{a4.6}) is equivalent to expression (%
\ref{a4.9}) 
\begin{equation}
F_{4}=-\frac{\hbar }{2(\rho +f^{s}j_{s})}\varepsilon _{iklm}[\partial
^{k}(j^{i}+f^{i}\rho )](j^{l}+f^{l}\rho )[\xi ^{m}-f^{m}(\xi ^{s}f_{s})]
\label{c9.1}
\end{equation}

To prove this statement, one sets $f^{0}=1,$ \ $f^{\alpha }=0$ in the
relation (\ref{c9.1}) and expands it 
\begin{eqnarray*}
F_{4} &=&-\frac{\hbar }{2(\rho +f^{s}j_{s})}\varepsilon _{iklm}[\partial
^{k}(j^{i}+f^{i}\rho )](j^{l}+f^{l}\rho )[\xi ^{m}-f^{m}(\xi ^{s}f_{s})] \\
&=&-\frac{\hbar \xi ^{\mu }}{2(\rho +j^{0})}\left( \varepsilon _{i0l\mu
}\partial ^{0}(j^{i}+f^{i}\rho )(j^{l}+f^{l}\rho )+\varepsilon _{i\beta l\mu
}\partial ^{\beta }(j^{i}+f^{i}\rho )(j^{l}+f^{l}\rho )\right) \\
&=&-\frac{\hbar \xi ^{\mu }}{2(\rho +j^{0})}\left( \varepsilon _{\alpha
0l\mu }\partial _{0}j^{\alpha }j^{\beta }+\varepsilon _{0\beta \alpha \mu
}\partial ^{\beta }(j^{0}+\rho )j^{\alpha }+\varepsilon _{\alpha \beta 0\mu
}\partial ^{\beta }j^{\alpha }(j^{0}+\rho )\right)
\end{eqnarray*}
One substitutes the expression of $j^{i}$ via variables ${\bf v,}$ $\eta $ 
\begin{equation}
j^{0}=\rho \cosh \eta ,\qquad j^{\alpha }=\rho \sinh \eta v^{\alpha }
\label{c9.2}
\end{equation}
in this expression and obtains 
\begin{eqnarray*}
F_{4} &=&-\frac{\hbar \xi ^{\mu }\rho }{2(1+\cosh \eta )}\sinh ^{2}\eta
\varepsilon _{\alpha 0\beta \mu }\partial _{0}v^{\alpha }v^{\beta }-\frac{%
\hbar \xi ^{\mu }}{2}\varepsilon _{\alpha \beta 0\mu }\partial _{\beta
}\left( \rho \sinh \eta v^{\alpha }\right) \\
&&-\frac{\hbar \xi ^{\mu }\sinh \eta }{2(1+\cosh \eta )}\varepsilon _{0\beta
\alpha \mu }\partial _{\beta }\left( \rho (\cosh \eta +1)\right) v^{\alpha }
\\
&=&\frac{\hbar \xi ^{\mu }}{2}\frac{\rho \sinh ^{2}\eta }{(1+\cosh \eta )}%
\varepsilon _{0\alpha \beta \mu }\partial _{0}v^{\alpha }v^{\beta }-\frac{%
\hbar \xi ^{\mu }}{2}\rho \varepsilon _{0\alpha \beta \mu }\partial _{\beta
}\left( \sinh \eta v^{\alpha }\right) \\
&&+\frac{\hbar \xi ^{\mu }}{2}\frac{\rho \sinh \eta }{(1+\cosh \eta )}%
\varepsilon _{0\alpha \beta \mu }\partial _{\beta }(\cosh \eta +1)v^{\alpha }
\\
&=&-\frac{\hbar \xi ^{\mu }}{2}\rho \varepsilon _{0\alpha \beta \mu }\left( -%
\frac{\sinh ^{2}\eta }{(1+\cosh \eta )}\partial _{0}v^{\alpha }v^{\beta
}+\partial _{\beta }\left( \sinh \eta v^{\alpha }\right) \right) \\
&&+\frac{\hbar \xi ^{\mu }}{2}\rho \varepsilon _{0\alpha \beta \mu }\frac{%
\sinh \eta }{(1+\cosh \eta )}\partial _{\beta }(\cosh \eta +1)v^{\alpha } \\
&=&-\frac{\hbar \xi ^{\mu }}{2}\rho \varepsilon _{0\alpha \beta \mu }\left(
-2\sinh ^{2}\frac{\eta }{2}\partial _{0}v^{\alpha }v^{\beta }+\cosh \eta
v^{\alpha }\partial _{\beta }\eta +\sinh \eta \partial _{\beta }v^{\alpha
}\right) \\
&&+\frac{\hbar \xi ^{\mu }}{2}\rho \varepsilon _{0\alpha \beta \mu }\frac{%
\sinh ^{2}\eta }{(1+\cosh \eta )}v^{\alpha }\partial _{\beta }\eta \\
&=&-\frac{\hbar \xi ^{\mu }}{2}\rho \varepsilon _{0\alpha \beta \mu }\left(
-2\sinh ^{2}\frac{\eta }{2}\partial _{0}v^{\alpha }v^{\beta }+\sinh \eta
\partial _{\beta }v^{\alpha }\right) \\
&&+\frac{\hbar \xi ^{\mu }}{2}\rho \varepsilon _{0\alpha \beta \mu }\frac{%
\sinh ^{2}\eta -\cosh \eta -\cosh ^{2}\eta }{(1+\cosh \eta )}v^{\alpha
}\partial _{\beta }\eta \\
&=&-\frac{\hbar \xi ^{\mu }}{2}\rho \varepsilon _{0\alpha \beta \mu }\left(
-2\sinh ^{2}\frac{\eta }{2}\partial _{0}v^{\alpha }v^{\beta }+\sinh \eta
\partial _{\beta }v^{\alpha }+v^{\alpha }\partial _{\beta }\eta \right)
\end{eqnarray*}

The obtained relation coincides with the expression (\ref{a4.6}) for $F_{4},$
that proves correctness of expression (\ref{a4.9}).

\section{Transformation of equation for variable ${\bf \protect\xi }$}

Let us transform equation (\ref{b6.5}) 
\begin{equation}
{\bf \xi }\times \left( -\dot{{\bf \xi }}\times {\bf z}+{\frac{({\bf z}\dot{%
{\bf \xi }})}{2(1+{\bf z\ \xi })}}{\bf \xi }\times {\bf z}+{\frac{{\bf \xi }(%
\dot{{\bf \xi }}\times {\bf z})}{2(1+{\bf z\ \xi })}}{\bf z}-{\frac{(1+{\bf %
z\ \xi })}{2}}{\bf b}\right) =0  \label{f9.1}
\end{equation}
keeping in mind that ${\bf \xi }^{2}=1$ and ${\bf z}^{2}=1$. Two middle
terms could be represented as the double vector product 
\begin{equation}
{\bf \xi }\times \left( -\dot{{\bf \xi }}\times {\bf z}+{\frac{1}{2(1+{\bf %
z\ \xi })}}\left( {\bf \dot{\xi}}\times \left( {({\bf \xi }\times {\bf z}%
)\times }{\bf z}\right) \right) -{\frac{(1+{\bf z\ \xi })}{2}}{\bf b}\right)
=0  \label{f9.2}
\end{equation}
Or in the form 
\begin{equation}
{\bf \xi }\times \left( \dot{{\bf \xi }}\times \left( -{\bf z+}{\frac{\left( 
{\bf z\xi }\right) {\bf z-\xi }}{2(1+{\bf z\ \xi })}}\right) -{\frac{(1+{\bf %
z\ \xi })}{2}}{\bf b}\right) =0  \label{f9.3}
\end{equation}
Now calculating double vector products and taking into account that ${\bf %
\xi \dot{\xi}}=0$, one obtains 
\begin{equation}
-\dot{{\bf \xi }}-\left( {\bf \xi }\times {\bf b}\right) =0  \label{f9.5}
\end{equation}


\newpage

\begin{thebibliography}{99}
\bibitem{R98}  Yu. A. Rylov, 
{\it Found. Phys.} {\bf 28, }245, (1998).

\bibitem{C57}  A. Clebsch, 
{\it J. reine angew. Math.} {\bf 54 }, 293-312 (1857).

\bibitem{C59}  A. Clebsch, 
{\it J. reine angew. Math.} {\bf 56 }, 1-10, (1859).

\bibitem{R99}  Yu. A. Rylov, 
{\it J. Math. Phys.}, {\bf 40,} 256-278, (1999).

\bibitem{B52}  D. Bohm, {\it Phys. Rev., }{\bf 85}, 166, (1952);{\it \ }{\bf %
85}, 180, (1952).

\bibitem{M26}  E. Madelung, {\it Z.\ Phys. }{\bf 40}, 322, (1926).

\bibitem{B26}  L. de Broglie, {\it Comptes Rendus, }{\bf 183}, 447, (1926).

\bibitem{R91}  Yu.A.~Rylov, 
{\it J. Math. Phys.} {\bf 32}, 2092-2098, (1991).

\bibitem{R995}  Yu. A.~Rylov, 
in {\it Chaos: the interplay between stochastic and deterministic behaviour}%
, eds. P. Garbachewski, M. Wolf, A. Weron, (Karpacz'95 Proc. Lecture Notes
in Physics. Springer, Berlin), {\bf 457}, 523-529, (1995)

\bibitem{D58}  P. A. M. Dirac, {\it Principles of Quantum Mechanics,} 4th
ed. Oxford, 1958.

\bibitem{FW50}  L. L. Foldy, and S. A. Wouthuysen, {\it Phys. Rev., }{\bf 78}%
, 29, (1950).

\bibitem{S930}  E. Schr\"odinger, {\it Sitzungsber. Preuss. Wiss. Phys.
Math. Kl.} {\bf 24}, 418, (1930).

\bibitem{B84}  A. O. Barut, N. Zanghi, {\it Phys. Rev. Lett}. {\bf 52},
2009, (1984).

\bibitem{BB81}  A. O. Barut, A. J. Bracken, {\it Phys. Rev}. {\bf D23},
2454, (1981).

\bibitem{A81}  J. C. Aron, {\it Found. Phys.} {\bf 11}, 863, (1981).

\bibitem{H90}  D. Hestenes, {\it Found. Phys.} {\bf 20}, 45, (1990).

\bibitem{RV93}  W. A. Rodriges Jr., J. Vaz Jr. {\it Phys. Lett.} {\bf B318},
623, (1993).

\bibitem{A67}  J. L. Anderson, {\it Principles of relativity physics}.
Academic Press, New-York, 1967, pp 75-88.

\bibitem{S30}  F. Sauter, {\it Zs. Phys. }{\bf 63,} 803, (1930), {\bf 64},
295, (1930).

\bibitem{S51}  A. Sommerfeld, {\it Atombau and Spektrallinien.} bd.2,
Braunschweig, 1951.

\bibitem{S61}  S. S. Schweber, {\it An Introduction to Relativistic Quantum
Field Theory. }New York, 1961, chp. 4, sec.3.

\bibitem{R999}  Yu. A. Rylov, 
{\it Adv. Appl. Cliff. Algebras}, {\bf 9}, 177, (1999).
\end{thebibliography}
\end{document}